# Technological Learning and Innovation Gestation Lags at the Frontier of Science: from CERN Procurement to Patents

Andrea Bastianin[a], Paolo Castelnovo[b,*], Massimo Florio[b] and Anna Giunta[c]

10 April 2019

**Abstract.** This paper contributes to the literature on the impact of Big Science Centres on technological innovation. We exploit a unique dataset with information on CERN's procurement orders to study the collaborative innovation process between CERN and its industrial partners. After a qualitative discussion of case studies, survival and count data models are estimated; the impact of CERN procurement on suppliers' innovation is captured by the number of patent applications. The fact that firms in our sample received their first order over a long-time span (1995-2008) delivers a natural partition of industrial partners into "suppliers" and "not yet suppliers". This allows to estimate the impact of CERN on the hazard to file a patent for the first time and on the number of patent applications, as well as the time needed for these effects to show up. We find that a "CERN effect" does exist: being an industrial partner of CERN is associated with an increase in the hazard to file a patent for the first time and in the number of patent applications. These effects require a significant "gestation lag" in the range of five to eight years, pointing to a relatively slow process of absorption of new ideas.

[a] University of Milan-Bicocca and CefES.
[b] University of Milan.
[c] Roma Tre University and Rossi-Doria Centre.

[*] *Corresponding author*: Department of Economics, Management and Quantitative Methods, University of Milan. Via Conservatorio, 7. 20122 - Milan – Italy.
Email: paolo.castelnovo@unimi.it

# 1. Introduction

The term "Big Science" identifies the style of scientific analysis characterizing much research in physics, astronomy and biology after World War II (Dennis, 2017). Big Science Centers (BSC) typically rely on large-scale instruments and research infrastructures, have a long list of participating institutions, attract generous funding from governments or international agencies and involve numerous industrial partners to develop the technologies required for research purposes. This last characteristic makes BSC a natural testing ground for assessing the effectiveness of Public Procurement for Innovation (PPI) as driver of firms' innovation output. In fact, public procurement through BSC is a form of PPI (see e.g. Aschhoff and Sofka, 2009[1]; Edquist and Hommen, 2000 for an overview of innovation policies).

Providing empirical evidence that procurement through BSC affects the innovation output of firms is key for governments contributing to their budgets. In fact, BSC are financed not only for the promise of significant scientific discoveries, but also in the hope that a radical technological innovation - yielding spillovers in different application domains (Dahlin and Behrens, 2005) - might arise as a side-effect while attempting to advance human knowledge (Hallonsten, 2014). The search for breakthrough or general-purpose technologies is an important target of innovation policies (Edler and Fagerberg, 2017). Being sources of aggregate productivity growth (Crépon et al., 1998), R&D and innovation represent key drivers of short - and medium-term business cycle fluctuations (Basu et al., 2006; Comin and Gertler, 2006; Kung and Schmid, 2015) and forces affecting long-run economic growth (Bresnahan and Trajtenberg, 1995; Schaefer et al., 2014).

There are both opportunities and challenges for BSC industrial partners: the latter are required to deliver new products with technology specifications developed for scientific

---

[1] Aschhoff and Sofka (2009) compares four public policies to stimulate innovation: public procurement, regulation, R&D subsidies, and subsidies to universities and research institutions. They find a positive effect of public procurement on market opportunities, and no difference between public procurement and access to knowledge created by universities and research institutes. The other channels seem to be less effective

purposes and for which a market might not yet exist. Consequently, there are risks related to the specificity of the technology developed for the BSC that might not be immediately exploited in other markets.

On the other hand, beneficial effects on the medium - long-term innovation output of suppliers might also arise. In fact, innovation and entrepreneurship might emerge from BSC through three main channels: (*i*) technological breakthroughs leading to start-ups and spin-offs involving BSC employees; (*ii*) inventions and patents filed by researchers at universities collaborating with the BSC; (*iii*) new business opportunities for BSC industrial partners. In some cases, these three channels may overlap (e.g. suppliers which employ former BSC staff or that are spin-offs of universities involved in experiments at the BSC).

In this paper we focus on the third channel, that is the effect of procurement through BSC on the output of firms' innovation activity as measured by patents.[2] We exploit a unique dataset with information about suppliers of the European Organization for Nuclear Research (CERN) as well as 28 case studies collected with face-to-face interviews to CERN industrial partners (Sirtori et al., 2019). CERN features a unique combination of both PPI and scientific research. The novelty of the scientific equipment needed at CERN to advance the frontier of research in physics is such that it is not obvious that its industrial partners will be able to benefit from such a collaboration. CERN suppliers face entirely new technological challenges that often require to advance the technological frontier, to acquire new scientific knowledge through R&D and develop radical innovations. The probability that firms can profit from their collaboration with CERN in the medium - long-term depends on their ability to enrich their absorptive capacity over time. Absorptive capacity captures firms' ability to absorb external knowledge and hence to benefit from the interaction process with a BSC (Cohen and Levinthal, 1990). Yet, the enrichment over time of firms' absorptive capacity is a necessary but not sufficient condition for an industrial partner to catalyse benefits originating from the procurement relationship with CERN. In particular, whether the experience gained through the interaction with CERN can be translated into innovations with a sizeable economic value remains an open issue.

---

[2] See e.g. Crépon et al. (1998), Aghion et al. (2013) and Jia et al. (2019) for the use of patents as a proxy innovation output.

The aim of the paper is twofold. First, we estimate survival and count data models to determine the impact of CERN procurement on the hazard to patent for the first time and on the number of patents filed by CERN suppliers. The fact that firms in our sample received the first order from CERN over a long-time span (1995-2008) delivers a natural partition of industrial partners into "suppliers" and "not yet suppliers". The time variation in the firms' status can thus be exploited to trace the impact of CERN on patenting activity. Second, besides evaluating the impact of CERN procurement on its industrial partners' patenting activity, we also investigate the average time-lag that separates CERN procurement from the filing date of a patent application. Specifically, we focus on the time that separates the start of the procurement relationship with CERN to the patent application date - if any - that proxies the end of the R&D project. In the parlance of Pakes and Shankerman (1984) this time lag is known as "gestation lag".[3]

The question pertaining to the time lag is relevant for policy makers who decide how much to invest in BSC based on member countries' investment returns. As shown by previous literature, the quantification of both the social and private rates of return to research rely on estimates of such time-lags. Mansfield (1968) used the time lags between the investment in academic research and the industrial utilization of their findings in the computation of the social rate of return to academic research. Pakes and Schankerman (1984) showed that the time lag between the deployment of research resources and the beginning of the stream of private revenues from their commercial applications heavily influences the private rate of return to research. Griliches (1979) surveyed the econometric literature concerned with the estimation of the return to R&D and highlighted that the comparisons across empirical analyses is difficult due to the variability of the assumption related with time lag R&D effects.

---

[3] These authors define the "total R&D lag" as the average time between the beginning of R&D expenditure and the start of the associated revenue stream. They further decompose the "total R&D lag" into the lag between project inception and completion - the "gestation lag" - and the time from project completion to commercial application - the "application lag".

We contribute to the literature on the impact of BSC on technological innovation in two ways. First, we advance knowledge on the innovative outcomes generated by BSC, focusing on patents. Although, recently, the economic literature has increasingly focused its attention on the impact of BSC procurement on industrial partners, this remains an under-researched area. Secondly, we shed new light on the time required for the BSC suppliers to "absorb" the technological content of the order and translate it into a potentially marketable innovation. To the best of our knowledge, this is the first paper that tackles such an issue in the case of BSC. The topic is investigated combining both qualitative and quantitative research techniques. Case studies allow us to take a closer look at different features of the interaction process between CERN and its suppliers and exploit "soft information" relevant to the design of our research hypotheses and econometric analysis.

We attain two main results: first, our investigation shows that a "CERN effect" does exist and is associated with an increase in both the hazard to file a patent for the first time and the number of patent applications; second, such effect is statistically significant only with a delay of some years from the beginning of the procurement relationship. The existence of a time-lag (i.e. 5 – 8 years) between CERN procurement and innovation confirms some evidence derived from our case-studies by signaling that learning from technologies at the frontier of science and translating such new knowledge into commercial applications is a medium-run process. This points to an absorption mechanism requiring protracted learning and adaptation. As underlined by Hameri and Vuola (1996: 131), while incremental innovation may easily access the market, "*the incubation times for revenues from a new* *technological application range from several years up to a decade, depending on the novelty of the solution*".

The remainder of the paper is organized as follows. Section 2 briefly presents CERN procurement practices, while Section 2.1 reviews the relevant literature on the impact of CERN procurement on industrial partners' sales, innovation, productivity and profitability. In Section 2.2 we go through case studies to uncover interaction process features and illustrate collaborative mechanisms between CERN and its industrial partners; afterwards, in Section 2.3, we illustrate opportunities and challenges arising for BSC suppliers and their consequences for learning and innovation. Against this background, in Section 2.4 we outline our research questions and hypotheses. Section 3 presents the data and the empirical strategy; Section 4 illustrates the results and several robustness checks, while Section 5 concludes. An Appendix with further details on the data and additional results completes the paper.

**2. CERN: background and procurement policies**

CERN - founded after World War II with the aim of studying the basic constituents of matter -operates the largest particle physics laboratory in the world and is a leading example of BSC. CERN research is publicly funded by 23 Member States according to their Gross Domestic Product. In turn, Member States expect an industrial return proportional to their contribution to CERN annual budget. CERN's experimental collaborations[4] involve over 17,500 people belonging to about 1,500 institutes from all over the world (CERN, 2018). The construction cost of its main research infrastructure - the Large Hadron Collider (LHC) – was more than 4 billion of Swiss Francs (CERN, 2019b). During the 1995-2015 period CERN collaborated with over 4,200 firms (CERN, 2019a), hence representing an ideal testing ground to investigate whether BSC generate industrial knowledge spillovers.

---

[4] There are seven experiments at the Large Hadron Collider (LHC) each addressing different research issues in particle physics. These are ATLAS, CMS, ALICE, LHCb, TOTEM, LHCf, and MoEDAL.

Several technologies, some of which completely innovative, were developed to build and operate the LHC. CERN procurement contracts have led to technological advances in superconductivity, cryogenics, electromagnets, ultra-high vacuum, distributed computing, rad-resistance materials, and fast electronics (Evans, 2009; Giudice, 2010). It follows that many of CERN procurement contracts require cutting-edge technologies and radical innovations combined with an intense collaboration with its industrial suppliers. CERN suppliers are exposed "*to a highly diverse knowledge environment*" (Autio et al., 2004: 110) which could positively impact on their expected future innovativeness, productivity and profitability. Procurement through CERN is thus not a form of "general public procurement" (i.e. buying off-the-shelf products), but rather a form of PPI. PPI shapes the demand environment and the economic landscape in which suppliers operate and can also impact on the innovation output of BSC's industrial partners (Uyarra and Flanagan, 2010).

**2.1 Brief literature review on the economic effects of CERN procurement**

The effects of CERN procurement on the innovation output of its industrial partners has been investigated with a variety of methodologies ranging from case studies to econometric analyses: the consensus view emerging from this strand of the literature points to the existence of a positive association between CERN procurement and firms' innovation output.

A survey of CERN suppliers showed that collaboration with CERN contributed -along with other factors - to product innovation and new R&D (Autio, 2014). Castelnovo et al (2018) relied on a simultaneous equation model to show that after becoming CERN industrial partners, firms generally experienced a rise in R&D, patents, productivity, and profits. Florio et al. (2018) showed that CERN procurement significantly affects suppliers' innovative performance when a relational governance is in place through cooperative relations (i.e.

exchanges implying that CERN and suppliers regularly cooperate to deal with complex information that is not easily transmitted or learned).

Based on evidence from 14 Swedish firms, Aberg and Bengston (2015) pointed out that firms' product and process innovation occurs mostly when a development project is in place, (i.e. when CERN invites firms to participate in developing products that cannot be bought off-the-shelf). Vuola and Hameri (2006: 3) relied on nine in-depth case studies and concluded that CERN is "*a most fertile ground to enable and boost industrial innovation*". Autio et al. (2014), based on three in depth case-studies, found innovation benefits accruing to the firms involved by CERN through prototypes' mocking up and experimentation. Closely related contributions are those by Amaldi (2012), Nielsen and Anelli (2016), and Battistoni et al., (2016) highlighting that thanks to the collaboration with CERN several firms were subsequently able to develop new products for customers in other markets.

While, as already mentioned the consensus is that there is a positive economic impact of CERN procurement on its suppliers, there is scant evidence on the challenges, risks, and the "gestation lag" of the innovation process related with procurement through BSCs.

**2.2 CERN procurement: features and challenges.**

A very recent survey – co-authored by one of the authors of this paper – collected 28 case studies with in-depth face-to-face interviews to representatives of CERN's suppliers that received at least one order since 1995 (Sirtori et al., 2019). See Appendix A1 for details about the companies involved in the analysis.

The orders are related either to the construction and upgrade of the LHC or to other CERN research programmes. We draw from these case studies to illustrate three special features that characterize the relation between CERN and its industrial partners: (*i*) transfers of human capital; (*ii*) transfers of experience; (*iii*) openness of knowledge.

*Transfer of human capital.* Mutual trust and the exchange of knowledge between CERN and firms is enhanced by the understanding of scientists and engineers to be part of the same experimental context, even when they play different roles in the procurement relation (see Tuertscher, 2014). Human capital flows from CERN to the industrial partner. For example, the technical director and co-founder of a Spanish firm that delivered components for the quadrupole magnets of the LHC, is a former CERN senior engineer. An Italian firm, specialized in electronic devices, was founded by engineers working in the 1970s at the Italian Institute for Nuclear Research (INFN) and CERN. A Spanish micro-firm with just 10 employees was created by former doctoral students involved in an experiment at CERN, and later from their collaboration in radiation detection at three research institutes: the Barcelona Institute of Microelectronics (IMB-CNM) and the Institute of Particle Physics in Valencia and the University of Liverpool.[5] Without such exchange of human capital, trust and flows of knowledge would be much more difficult.

*Transfer of experience.* Experience gained through procurement is transferable between BSC and beyond. There is a contagion effect within the extremely selective market of hi-tech procurement for BSC based on reputation earned at CERN. For instance, an Austrian software company, providing supervisory control and data acquisition systems, reported benefits arising from the fact that when CERN scientists returned to their national institutions, they spread the word about the functionality and performance of their software. This helped to acquire new customers including leading centers for hadron-therapy of cancer in Austria and Italy, and with the ITER fusion experiment in France. An Italian company, after providing cryostats for CERN, was able to enter a collaboration with the ITER fusion project. Similarly, a German company reported to collaborate not only with CERN but also with other BSC such as DESY and the GSI Helmholtz Centre for Heavy Ion Research. Even a branch of a big multinational company, after developing new products and customizing existing ones (i.e. providing insulation and protection equipment) for CERN, was then able to exploit such products in other markets. Reputation earned with a BSC is recognized elsewhere because scientists and engineers share a common background and understand how challenging it is to be involved in procurement at the frontier of science.

---

[5] The CERN Alumni website (https://alumni.cern) reports many other similar examples.

*Openness of knowledge*. The BSC environment creates an unusual circulation of knowledge. An interview with representatives of a multinational supplier revealed that: "*by interacting frequently with CERN staff, the company has also improved its organizational capabilities. In particular, it has learned the value of working closely with its customers – cooperating on some activities, including on product design, but also visiting their sites in person, to see where the procured products are actually used. Such field work is often impossible in industry projects, because of the strict confidentiality. Nevertheless, it is essential to ensure clear communication and to deliver precisely what the customer requires*" (Sirtori et al., 2019, p. 61).

An Austrian software company reported that their relationship with CERN went well beyond the usual customer–supplier relationship, making it more similar to a partnership. The company has subsequently joined "CERN Openlab" (https://openlab.cern), a public-private partnership to explore new ideas for future R&D projects. A software multinational firm, acknowledges that CERN played the role of lead user, thus improving the final quality of its products. These examples point to a knowledge environment which is more open than the usual business context.

Overall, most of the representatives interviewed declared that the relation with CERN had "*high effect*" both in improving the technical know-how (75%) and the reputation (64%) of firms. However, the interviews also highlight that the impact of CERN on sales or new customers' acquisition is not as strong as its effects on firms' technical know-how and reputation. This difference is interesting and suggests that some critical issues are often overlooked by the previous literature. It thus seems naive to conclude that procurement through BSC immediately translates into marketable innovations.

**2.3 Challenges and risks**

Some of the suppliers interviewed by Sirtori et al. (2019) reported that the collaboration with CERN has been challenging, costly, risky, and was not necessarily financially profitable in the short term. A superconductor manufacturer (part of a German company with 6,000 employees active in life science and analytical systems) supplied super-conducting cables for three generations of CERN particle accelerators (LEP in the 1980s, the LHC in the 1990s, HL-LHC currently), and is now involved in the conceptual design of the Future Circular Collider. Despite this long experience, the firm's representatives reported that the company had to manage high uncertainties in time and costs associated with some new technologies requested by CERN. Such a high degree of uncertainty required frequent interactions with CERN scientists and engineers. This is often the unavoidable side-effect of the learning benefits from BSC. Orders for off-the-shelf products are entirely specified and uncertainty is minimal, however contracts for highly customized products require several adjustments. Overall, CERN's orders for this company were not particularly profitable and, after their delivery, the company needed to downsize the number of its employees to adjust for new market conditions. Another Italian company confirmed that investing in R&D for a product whose precise technical requirements are not available, implies highly uncertain manufacturing costs and revenues. This has led to financial losses for this company.

A Swiss based company specialized in vacuum creation and gas-management has collaborated with CERN since the 1960s. Some of the orders from CERN for vacuum technologies were costly and risky for the company; this explains why the representatives of this firm declared that now they carefully evaluate whether to participate in a tender for CERN procurement contracts. The volume of the order and the schedule for its delivery are key elements in this decision and risks need to be compared with learning benefits and potential profits in other markets. An Italian company that has collaborated with CERN and other BSC reported that orders that are less technically challenging often yield higher profits, while learning opportunities are higher when the level of customization is higher.

Another challenge is related to subcontracting. For example, a Portuguese company specializing in energy production storage equipment had to involve subcontractors and to collaborate with other suppliers to fully meet CERN's stringent technological requirements. This required intensive interactions with CERN, new investment, training of its own staff to test new tools and solutions, selection and coordination with other firms. The need to involve subcontractors or to act as part of consortium adds complexity in the governance of the procurement relation and ultimately increases the uncertainty.

To provide further examples of why uncertainties and costs arise for suppliers, one may consider the following case stuies. An Austrian civil engineering company was required by CERN to keep two caverns where equipment had to be located as close as possible to minimize the length of optical cables needed for data transfer. While this choice is optimal from an Information Technology perspective, it revealed to be a particularly challenging task for civil engineers (i.e. less rock between caverns reduces the support of the structure). The company had to design new solutions for reinforcement but at a later stage it discovered that the rock itself had many soft layers and hence a sophisticated and costly groundwater control-system was necessary to prevent unexpected flooding during excavation. A second example is provided by a German company that had to develop an entirely new high-temperature furnace. After intensive and frequent interaction between the CERN's staff and the company, the furnace was designed, manufactured and delivered. This took 53 weeks of dedicated effort. A Dutch company specialized in aluminum technologies was required to design and manufacture a support structure for the LHC cryostats. There was no previous experience to comply with the required standards and the company had to employ special manufacturing equipment and techniques. Moreover, it also had to carry out R&D to understand how to treat aluminum at extremely low temperatures (LHC operates at 1.9 degrees Kelvin, lower than the vacuum in outer space). This extreme requirement needed additional costly investments for the firm.

These examples highlight some perhaps less known side-effects of technological learning through BSC procurement. It is an expensive process that requires frequent interactions with the BSC, additional R&D, specific fixed investment, training costs, use of subcontractors and financial risk. Moreover, customized solutions may not prove profitable in other markets. The high degree of uncertainty about how, when and if all these costs will be balanced by additional gains has inspired our research hypotheses about the gestation lags from procurement to innovation output.

## 2.4 Research hypotheses

Findings from our case studies suggest that collaborating with BSC involves a risk-return trade-off. On the one hand, BSC industrial partners might experience benefits related to technological learning and reputational effects that can potentially be exploited in their relations with other BSC or customers in different markets. On the other hand, entering in a contract with a BSC is not entirely without risk for suppliers. In fact, procurement contracts for technologies required to advance the frontier of science are often not profitable in the short-term. How benefits and costs will be balanced in the medium - long-term is uncertain. Firms engage in contracts with a BSC such as CERN not only for an immediate profit, but also with the expectation of future competitive advantages, which would require further steps.

This suggests that there might be a time lag that separates the beginning of the relation with the BSC and actual economic gains arising from it.

In this paper we focus on the time distance that runs from the first procurement contract to its completion, as measured by the filing date of a patent, if any. While, as far as we know, there is no previous literature on BSC-specific "gestation lag", several papers investigated the time lag associated with the commercial exploitation of academic research. Mansfield (1991, 1998) reported that the mean time interval between the relevant academic research and the first commercial introduction of new products or processes is 6-7 years. The estimated time span between the appearance of academic research and its effect on productivity in the form of knowledge absorbed by an industry might be even longer, on average approximately 20 years (Adams, 1990). In the case of computer science and engineering – two cases that are especially relevant also for what concerns CERN – the time lag is 10 years (Adams, 1990). Heher (2006) showed that it can take up to 10 years for an institution, and 20 years nationally, to attain a positive rate of return from an investment in research and technology transfer. In the pharmaceutical industry the time-lag between research, development, and commercialization can reach up to 20 years (see Sternitzke, 2010 and Toole, 2012). In contrast, earlier analyses by Pakes and Griliches (1980), Hausman et al. (1984), Hall et al. (1984) reported that about a year is necessary for translating R&D expenditure into a patent application.

What happens in the BSC context? The following research hypotheses seem to be justified by the previous qualitative discussion and earlier literature:

H1. *Firms engaged in a procurement relation with a BSC - such as CERN - learn to develop new technologies through the co-development of designs related to procurement contracts. This learning process ultimately leads to an innovation output, against initial R&D expenditure and other costs faced by firms.*

> H2. *Given the idiosyncratic nature of the new developments arising from the procurement for a BSC, the subsequent potential adaptation and application to other customers takes significant time and effort to produce potentially marketable innovation output.*

The empirical proxy of innovation output we used to test H1 is the *number of patent applications* filed by CERN suppliers after the start of the procurement relation with CERN. Patents are widely used as a proxy of innovation output (see e.g. Crépon et al., 1998; Aghion et al., 2013; Jia et al., 2019), although they may underestimate the innovation effect of BSC procurement since many firms might not patent their innovations or use alternative forms of intellectual property protection (see e.g. Dziallas and Blind, 2019 for a critical overview of innovation indicators). Our empirical proxy for testing H2 is the *gestation lag*: namely, the time that separates the beginning of an R&D project and its completion. We assume that the beginning of the procurement relation with CERN marks the starting date of a specific R&D project, while the patent application date - if any - represents its conclusion. We thus want to empirically assess the likelihood that CERN suppliers are able to increase their patenting activity (H1) and how long is the gestation lag needed to absorb the new ideas and translate them into potentially marketable innovations (H2).

## 3. Data and Empirical strategy

### 3.1 Data

We have assembled a unique dataset collecting information on firms collaborating at the development of the world biggest research infrastructure: the LHC, and some related projects at CERN. The LHC project was approved by the CERN council in December 1994, while the LHC Conceptual Design Report, which detailed the architecture and operation of the LHC, was published in October 1995. The experiments at the LHC started in September 2008 and in July 2012 the discovery of the Higg's boson – for which François Englert and Peter Higgs were awarded with the 2013 Nobel Prize in physics - was announced to the public.

Our dataset summarizes information from three sources. First, the CERN Procurement and Industrial Services Group – that is in charge of coordinating all the supplies and services that the laboratory needs – maintains a database that we used to identify firms that over the 1995-2008 period received at least one LHC-related order above 10,000 CHF.[6] We exploited this database also for retrieving the date that marks the beginning of the procurement relationship, the "activity codes" used by CERN to classify purchases from its suppliers, the number and the total amount of LHC-related orders.

Second, we sourced balance sheet data for LHC suppliers from the ORBIS database maintained by Bureau van Dijk. We collected information on the geographical location, size and sector of activity (based on NACE 2 digits codes) of firms, as well as data on the amount of their intangible fixed assets. Firms' intangible fixed assets in the ORBIS database comprise all intangible assets such as formation expenses, research expenses, goodwill, development expenses and all other expenses with a long-term effect. Lastly, we retrieved the number of patents filed by LHC suppliers each year from the PATSTAT database.

The process of merging data from the CERN procurement database, ORBIS and PATSTAT left us with a panel dataset with 896 firms out of 1296 that have collaborated with CERN at the construction of the LHC. See Section 3.2 for further details on the sample design. For each year in the 1995-2008 period, Figure 1 shows the number of firms that have received their first LHC-related order. As we can see, except for 1995-96 and 2007-08, firms are almost uniformly distributed over time.

---

[6] The rationale for excluding low value orders is that they are unlikely to generate knowledge spillovers for firms. Moreover, notice that experimental collaborations, such as ATLAS and CMS, have some procurement autonomy, so their orders are not covered here, except when they are directly managed by CERN. For additional details see: http://procurement.web.cern.ch/procurement-strategy-and-policy

**Figure 1. Time distribution of LHC-related orders: 1995-2008**

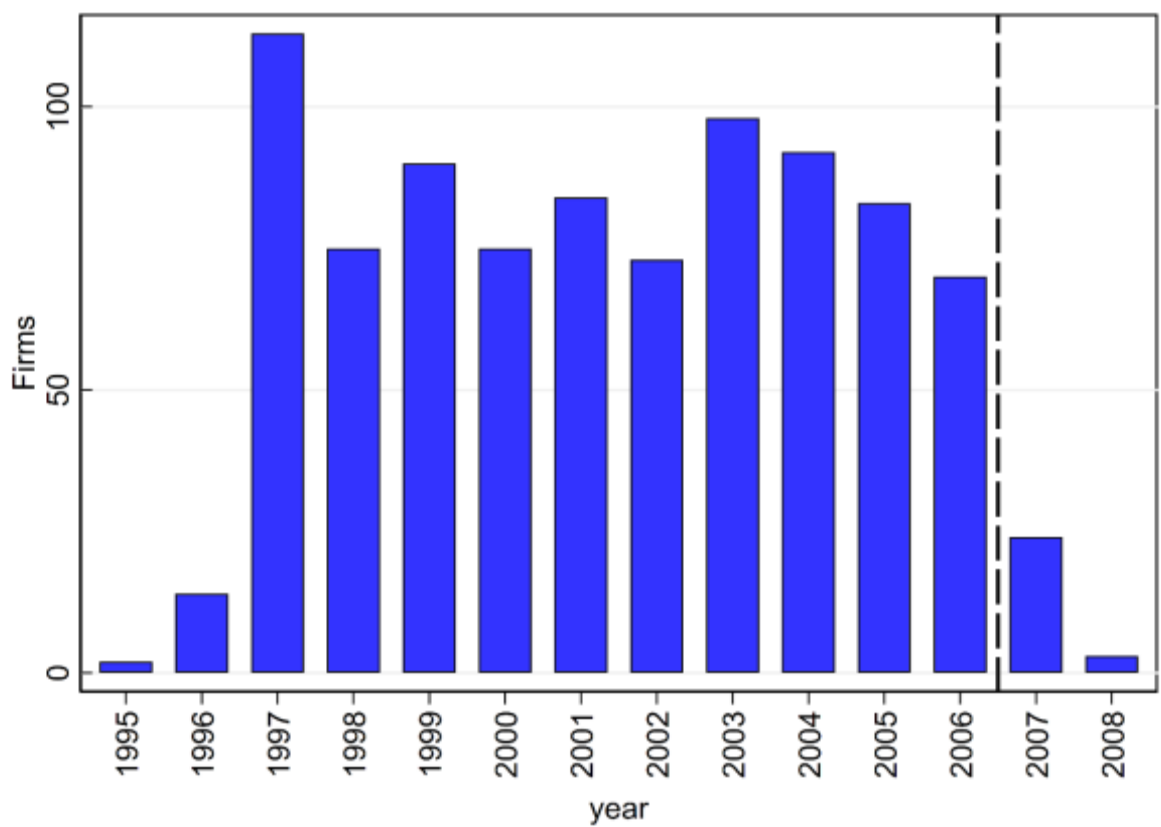

*Notes*: for each year in the 1995-2008 period, the figure shows the number of firms that have received their first LHC-related order. The end of the sample period used in the analysis is 2006 as highlighted by the vertical dashed line.

## 3.2 Empirical strategy

The empirical analysis is based on two steps. First, we rely on survival analysis to assess whether the collaboration with CERN has increased the hazard of filing a patent for the first time. Second, we estimate count data models to quantify the impact of CERN procurement on the number of patents filed by LHC suppliers. This can be accomplished because collaborating firms have received their first order from CERN over a long time-span that delivers a natural partition of statistical units into "suppliers" and "not-yet-suppliers". In both cases we also pay attention to the timing of the "CERN effect".

**Figure 2. Time distribution of LHC suppliers: 1993-2006**

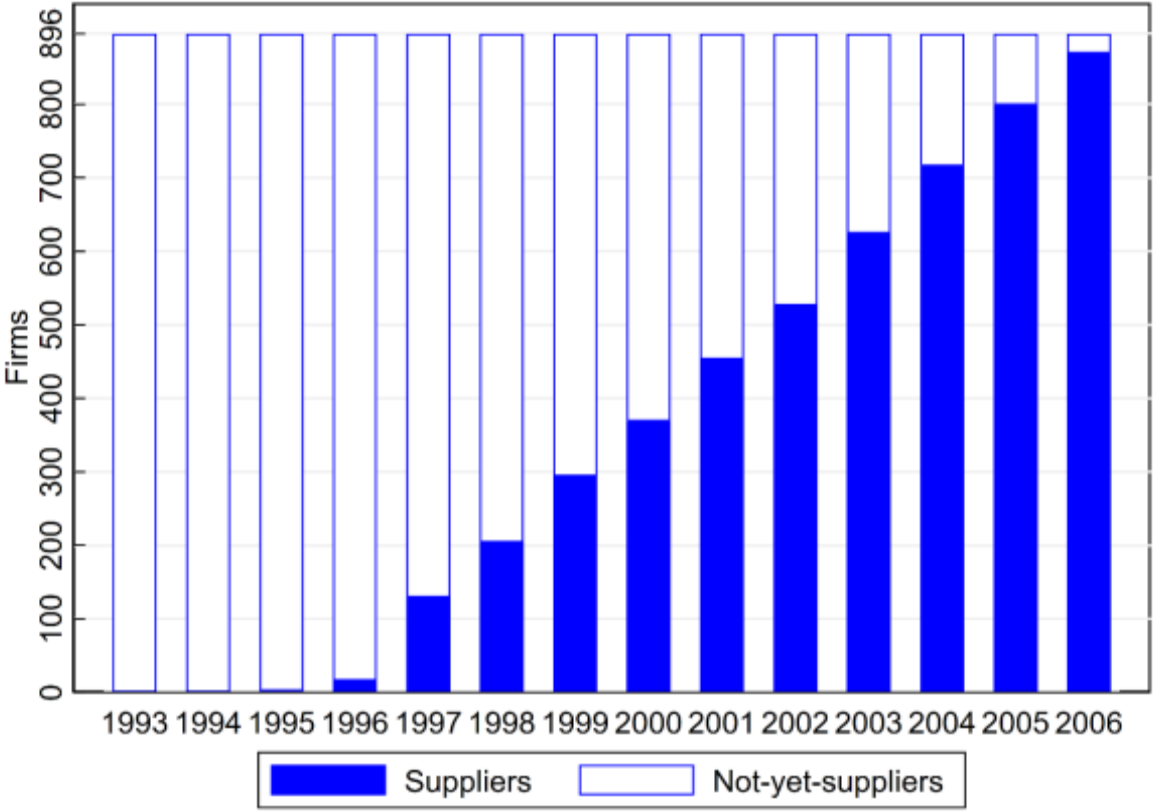

*Notes*: figure shows the transition of firms from "not-yet-supplier" (white bars) to "LHC supplier" (filled bars). We equate the transition event to the beginning of the procurement relationship.

The time variation of firms' status can thus be exploited in the econometric analysis to evaluate how CERN procurement has affected the patenting activity of firms. Notice that the LHC was constructed over the 1995-2008 period, but the empirical investigation relies on a panel dataset for the 1993-2006 period. The beginning of the observation period pre-dates the start of the construction of the LHC because we are interested in the timing of the "CERN effect". To be sure that such effect does not pre-date the start of the procurement relation, some of our empirical specifications include a set of dummy variables taking value one if the company will become a CERN supplier in one or two years. The coefficients on such dichotomous variables - that act as leads – should be never statistically distinguishable from zero in case of a casual interpretation of the "CERN effect" on firms' patent applications[7]. As shown in Figure 2, before 1995 none of the companies in our sample received an LHC-related order. Moreover, we decided to stop our follow-up period at year 2006 rather than 2008, since as of 2006 not all of the 896 firms had started their procurement relation with CERN. In fact, we have 869 firms changing their status from "not-yet-suppliers" to "LHC supplier". The remaining 27 entities are included as controls.[8]

### 3.2.1 Survival analysis

We rely on survival analysis to estimate the determinants of the hazard of patenting (Kiefer, 1988 for an overview of survival analysis in economics). Since we are interested in observing the transition of statistical units from the status of "not-patenting" firm to that of "patenting" firm, we excluded entities that have filed patents before 1995. This is a strict criterion that leaves us with 740 firms, out of 896, for which a balanced panel dataset is available. Among them, 158 (21.4%) have filed their first patent after the beginning of their collaboration with CERN.

We consider the Cox Proportional Hazard model (Cox, 1972) that specifies the hazard of patenting of firm $i$ in year $t$ as:

$$\lambda(t|\mathbf{X}_{isc,t}) = \lambda_0(t) exp\left(\gamma_k CERN_{i,t}^k + \mathbf{X}_{i,t}\boldsymbol{\beta} + \delta pct_{c,t} + \sum_s \alpha_s + \sum_c \alpha_c\right) \quad (1)$$

the hazard rate, $\lambda(.)$, is the instantaneous probability of applying for a patent in year $t$ for the first time for firms that have not yet filed a patent by year $t$. The baseline hazard, $\lambda_0(t)$, is a function of time alone. The exponential function ensures that $\lambda(.)$ is nonnegative and acts as a scale factor that makes the baseline hazard proportional to the vector of the covariates $\mathbf{X}_{isc,t}$. This vector includes all explanatory of the variables – including the dummies – appearing on the right-hand side of Equation 1.

---

[7] As shown in Section 4.2 our results are unaffected if the leads are omitted from the model.
[8] As explained in Sections 4.1 and 4.2, empirical analyses rely on samples of different size. The survival analysis includes 740 firms. As of 2006, 716 of them had changed their status from "not-yet-suppliers" to "LHC suppliers". Count data models are estimated on a sample of 100 firms. As of 2006, 99 had changed their status from "not-yet-suppliers" to "LHC suppliers".

The dummy variables $CERN_{i,t}^{k}$ is set to one if the procurement relation with CERN started *k* years ago. The contribution of country *c* to CERN budget in year *t* – expressed as percentage of the total contribution of Member States – is denoted as $pct_{c,t}$. This variable controls for country heterogeneity: it captures the fact that firms located in countries contributing more, might have a higher probability of receiving an order and hence to benefit from knowledge spillovers. This also because of the official CERN policy of balancing orders across Member States. Additional factors capturing unobservable country-specific heterogeneity are modelled with country dummy variables, $\alpha_c$. Similarly, we also include sector specific dummy variables whose aim is to allow the baseline hazard to vary across sectors.

The vector $\mathbf{X}_{i,t}$ includes firm-level control variables used to capture factor that might affect the absorptive capacity of firms (Cohen and Levinthal, 1990). In fact, following Autio et al. (2004) we posit that industrial learning effects of BSC are generated by their dyads with firms (i.e. a "dyad" is a relationship between organizations). Whether firms can exploit knowledge spillovers will depend both on the firms' absorptive capacity and on the absorptive capacity of the dyad (Lane and Lubatkin, 1998), namely the sum of relation-specific assets that facilitate both knowledge disclosure and knowledge communication within the dyad.

The effect of firm size on the hazard to file the first patent is captured by dichotomous variables classifying firms into small (i.e. the reference category), medium, large and very-large. This classification - provided in the ORBIS database - exploits information on the amount of total assets, operating revenues and the number of employees. Firm size is expected to be positively associated with the hazard to patent because large firms can exploit economies of scale, have access to a broader pool of highly qualified collaborators (Fernández-Olmos and Ramírez-Alesón, 2017) and have more financial resources to afford the costly process of patent application (see Block et al. 2015, Blind et al. 2006, and Leiponen and Byma, 2009, among the others).

We also include in $\mathbf{X}_{i,t}$ a dummy variable ($Hi\text{-}tech_i$) that exploits the technological intensity of the orders received by firms to classify them as hi- or lo-tech. Hi-tech firms (i.e. those that have received at least one order classified as hi-tech by CERN experts) might have higher absorptive capacity and be more capable of translating hi-tech orders from BSC into marketable innovations (Castelnovo et al., 2018; Hameri and Vuola, 1996; Edquist and Hommen, 2000). Details on the construction of this variables and robustness checks of results based on an alternative definition of firms' technological intensity are presented in the Appendix A2. Lastly, we consider the (logarithm of the) total amount of LHC orders received by firm $i$ ($Order_i$) as a proxy of the involvement (e.g. joint meetings over several years) and continuity of the procurement relationship with CERN (Åberg and Bengtson, 2015). In fact, as previously mentioned, long-lasting collaborations for hi-value orders often involve repeated interactions with CERN that might boost the learning effects of procurement (Autio et al., 2004; Florio et al., 2018).

### 3.2.2 Count data analysis

Since the aim of the survival analysis is assess whether LHC procurement is positively associated with an increase in the hazard to patent for the first time, we rely on the subsample of LHC suppliers that have never filed a patent before 1995. With count data models we want to quantify the incremental number of patents that firms filed after the beginning of their relationship with CERN, therefore we focus only on firms that have filed at least one patent since their incorporation. This leaves us with exactly 100 firms for which a panel dataset over the 1993-2006 period is available (i.e. a total of 1400 observations). The main driver of the reduction of the sample size is the inclusion of a proxy of R&D spending in our empirical analyses. Robustness checks involving changes in this sample design are discussed at the end of Section 4.2.

Because the number of yearly patent applications filed by firms is a count – that is, a non-negative integer-valued variable - with many zero and ones, we estimate both Poisson and Negative Binomial models. The expected number of new patents filed by firm *i* each year, $p_{i,t}$, can be written as follows:

$$E(p_{i,t}|\mathbf{X}_{isc,t}) = exp(\sum_k \gamma_k CERN_{i,t}^k + \mathbf{X}_{i,t}\boldsymbol{\beta} + \delta pct_{c,t} + \sum_s \alpha_s + \sum_c \alpha_c + \sum_t \alpha_t) \qquad (2)$$

Many explanatory variables in Equation (2), including some of the firm-level controls in the vector $\mathbf{X}_{i,t}$, have already been introduced in Section 3.2.1. Regressions now include a set of dichotomous variables ($CERN_{i,t}^k$) capturing the effects of CERN procurement on patenting over time.[9] We also add time dummies ($\alpha_t$) to control for common macroeconomic shocks hitting all firms in the sample. Macroeconomic conditions are expected to affect the profitability of firms, the business environment where they operate and the relationship between technology collaboration networks and innovation performance (Fernández-Olmos and Ramírez-Alesón, 2017). The logarithm of Intangible Fixed Asset[10] ($IFA_{i,t}$) proxies R&D expenditure (Chan et al., 2001; Leoncini et al., 2017; Marin, 2014). R&D expenditure is a key control variable for analysing patent activity (Hall et al., 1984; Hausman et al., 1984; Aghion et al. 2013; Gurmu and Pérez-Sebastián, 2008), but unfortunately the proxy we use features many missing observations. However, other proxies of R&D expenditure available in the ORBIS database feature an even higher number of missing observations. To include $IFA_{i,t}$, while maintaining the size of the panel dataset reasonably large, we excluded firms reporting more than 4 missing values.[11] Robustness checks concerning this issue are discussed in Section 4.2.

---

[9] A similar approach was proposed by Stevenson and Wolfers (2006) in a completely different domain of analysis.
[10] Intangible fixed assets recorded in the Orbis database include all intangible assets such as formation expenses, research expenses, goodwill, development expenses and all other expenses with a long-term effect.

Lastly, following Blundell et al. (1999), we use the mean pre-sample patent count (i.e. *Avg. $p_i$*, the average number of patent applications per year in the thirty years before 1993) to capture the firm's unobserved propensity to patent. This variable captures unobservable firm-specific fixed effects reflecting any permanent differences in the level of innovation output across firms which are independent of CERN procurement.

## 4. Results

This section presents the results of the empirical analysis based on survival and count data models. While in both cases we also discuss a wide array of robustness checks, for the sake of brevity, the corresponding tables are confined in Appendix A3.

### 4.1 Effects and timing of CERN procurement on the hazard of patenting

Estimates of ten different specifications of the Cox Proportional Hazard model - Equation (1) are reported in Table 1. Each column of Table 1 reports estimates of the coefficient on a distinct "CERN effect" variable: $CERN_{i,t}^k$ for $k = 0, 1, ..., 9$. Recall that $CERN_{i,t}^k$ is a dummy variable equal to one if the procurement relation started $k$ years ago. Therefore, the coefficient on $CERN_{i,t}^0$ measures the change in the hazard rate in the year of the first LHC-related order. Similarly, $CERN_{i,t}^1$ captures the variation of the hazard rate one year after the first order, and so on and so forth.

---

[11] To have a balanced panel dataset, missing observations have been substituted with zeros and a dummy variable taking value one for such observations is included in regressions.

**Table 1. Cox proportional hazard model for start of patenting activity**

| | (1) $k=0$ | (2) $k=1$ | (3) $k=2$ | (4) $k=3$ | (5) $k=4$ | (6) $k=5$ | (7) $k=6$ | (8) $k=7$ | (9) $k=8$ | (10) $k\geq 9$ |
|---|---|---|---|---|---|---|---|---|---|---|
| $CERN_{i,t}^k$ | 0.034 | 0.043 | 0.046 | 0.054** | 0.067*** | 0.080*** | 0.088*** | 0.071* | -0.009 | 0.034 |
| | (0.047) | (0.037) | (0.029) | (0.022) | (0.023) | (0.022) | (0.023) | (0.040) | (0.059) | (0.073) |
| $Hi\text{-}Tech_i$ | 0.133 | 0.131 | 0.132 | 0.130 | 0.130 | 0.135 | 0.143 | 0.147 | 0.140 | 0.143 |
| | (0.176) | (0.175) | (0.174) | (0.175) | (0.175) | (0.174) | (0.175) | (0.175) | (0.175) | (0.175) |
| $Order_i$ | 0.001 | -0.003 | -0.004 | -0.006 | -0.009 | -0.012 | -0.009 | -0.000 | 0.013 | 0.010 |
| | (0.068) | (0.066) | (0.064) | (0.061) | (0.062) | (0.062) | (0.061) | (0.057) | (0.058) | (0.058) |
| $Medium_i$ | 0.515* | 0.516* | 0.517* | 0.522* | 0.529* | 0.533* | 0.529* | 0.523* | 0.519* | 0.520* |
| | (0.273) | (0.273) | (0.272) | (0.272) | (0.272) | (0.274) | (0.274) | (0.275) | (0.279) | (0.278) |
| $Large_i$ | 1.164*** | 1.162*** | 1.162*** | 1.166*** | 1.173*** | 1.178*** | 1.170*** | 1.164*** | 1.172*** | 1.168*** |
| | (0.398) | (0.396) | (0.395) | (0.395) | (0.396) | (0.400) | (0.404) | (0.409) | (0.410) | (0.410) |
| $V\,Large_i$ | 2.055*** | 2.057*** | 2.062*** | 2.071*** | 2.090*** | 2.103*** | 2.096*** | 2.071*** | 2.050*** | 2.051*** |
| | (0.370) | (0.363) | (0.362) | (0.359) | (0.360) | (0.359) | (0.362) | (0.367) | (0.379) | (0.378) |
| $pct_{c,t}$ | 0.002 | 0.002 | 0.001 | 0.001 | 0.002 | 0.001 | 0.002 | 0.002 | 0.002 | 0.002 |
| | (0.004) | (0.004) | (0.004) | (0.004) | (0.004) | (0.004) | (0.004) | (0.004) | (0.004) | (0.004) |
| $\alpha_c$ | yes | Yes | Yes | Yes | yes | Yes | yes | yes | yes | yes |
| $\alpha_s$ | yes | Yes | Yes | Yes | yes | Yes | yes | yes | yes | yes |

*Notes*: country ($\alpha_c$) and sector ($\alpha_s$) fixed effects have been included in all the specifications. * p-value < 0.10, ** p-value < 0.05, *** p-value < 0.01. Cluster robust standard errors (i.e. cluster is the 2 digits NACE code) in parentheses. The table shows estimates of the coefficients in Equation (1) and not hazard ratios.

Table 1 shows that the association between the LHC procurement and the variation of the hazard to file a patent for the first time is characterized by an inverted U-shaped relation. In fact, estimated coefficients on $CERN_{i,t}^k$ are positive, thus confirming *H1*, but not significant for $k = 1,2,3$ while significant and monotonically increasing up to $k = 6$ and then they decrease, thus confirming *H2*. Figure 3 plots the estimates of the hazard ratio associated with the estimated coefficients shown in the first row of Table 1 and the corresponding 95% confidence interval. An estimated hazard ratio equal to one means lack of association, while an estimate greater than one suggests that CERN industrial partners face a higher "risk" of filing their first patent compared to "not-yet-suppliers". As shown in Figure 3, the 90% confidence bands for $CERN_{i,t}^k$ lies above one only for $3 \leq k \leq 6$.

This implies that the effect of LHC procurement on the hazard ratio to file a patent for the first time takes time to build and shows up with a lag of at least 3 years. Similarly, the fact that coefficients on $CERN_{i,t}^k$ for $k \geq 8$ are not statistically distinguishable from zero implies that the association between LHC procurement and changes in the hazard rate vanishes eight years after receiving the first order from CERN.

**Figure 3. Estimates of the hazard ratio for LHC suppliers**

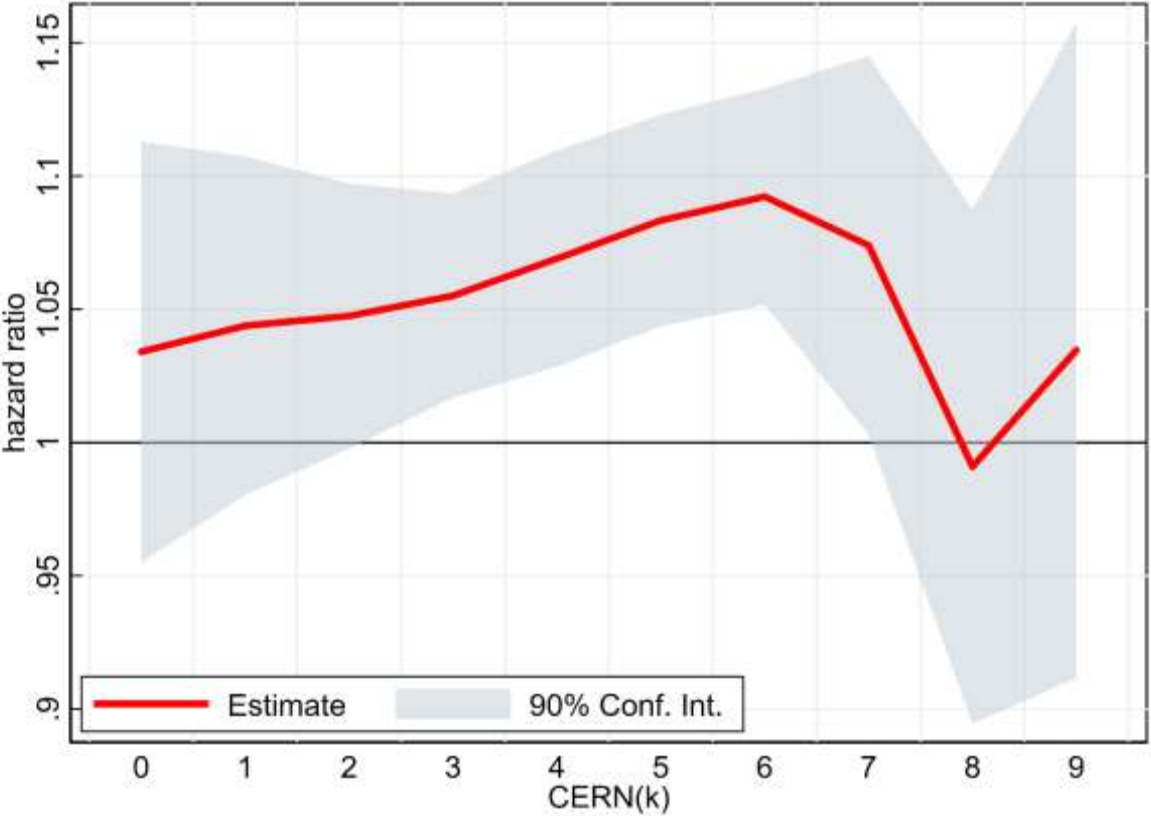

*Notes*: the figure plots the estimates of hazard ratio associated with the estimated coefficients shown in the first row of Table 1 and the corresponding 90% confidence interval. An estimated hazard ration greater equal to one means lack of association, while an estimate greater than one suggests that CERN industrial partners face a higher "risk" of filing their first patent compared to "not-yet-suppliers". CERN(*k*) is a dummy variable equal to one if the procurement relation started k years ago.

Moving to control variables, we see that independently of the specification, the estimated coefficients on size variables are positive and statistically significant at the 99% level for large and very-large firms. In line with the expectations, this suggests that, given all the other covariates, for large and very large firms the hazard to patent is higher than for small firms, which represent the reference category. All models in Table 1 also include country and sector fixed-effects. The Wald tests of the null hypothesis that country and sector fixed effects are jointly equal to zero suggest rejecting the null for both set of controls.

The remaining control variables are never statistically distinguishable from zero. More precisely, we see that, as expected, the hazard to patent is higher for hi-tech firms and increase with the percentage contribution to CERN budget of the country that hosts the firm. However, none of these variables are statistically distinguishable from zero. This applies also to the total amount of LHC orders received by firms; in fact, the coefficient on this variable switches sign across specification and there is no economically meaningful explanation for that.

The existence of a sizeable time lag between procurement from BSC and subsequent benefits for collaborating firms is in line with the findings in related strands of the literature. Griliches (1979), in his survey of econometric analyses of the R&D-productivity nexus, highlights that a bell-shaped lag structure connects firm R&D to changes in productivity; such a shape is due to the fact that it takes time before research can be fruitfully exploited by firms.

*Robustness checks.* The fact that the coefficient on the dichotomous variable used to classify statistical units into hi- and lo-tech firms is never statistically distinguishable from zero is somehow surprising, given our expectation that this variable might capture some aspects related with the absorptive capacity of firms. To verify that the large standard errors associated with these coefficients are not due to measurement error, we attempt to proxy the degree of technological intensity of firms with a continuous variable that captures the share of orders classified as hi-tech. Results in the Section A3.1 of the Appendix show that the findings of Table 1 are unaffected when using this alternative criterion to measure firms' technological intensity.

Similarly, when we replace the total amount of LHC orders with the (logarithm of the) orders count, as an alternative proxy of the involvement and continuity of the procurement relationship, our main results remain unchanged.

**4.2 Effects and timing of CERN procurement on patenting activity**

Figure 4 displays the number of patents per firm over "relative years" - denoted as *k*. These measure the time from the first LHC order. Therefore, *k* > 0 indicates that the first LHC order was received *k* years ago. The two horizontal lines in Figure 4 are sample averages for *k* ≤ 0 and *k* > 0. Before the beginning of the relationship with CERN (i.e. *k* ≤ 0), the sample average is 0.31 patents per firm, while after receiving the first LHC order (i.e. *k* > 0) the sample average rises to 0.93. A simple t-test for the equality of means allows to reject the null hypothesis and is therefore suggestive of the existence of a "CERN effect" on firms' innovation output.

The existence of a "CERN effect" is further investigated in Table 2, where we regress the number of patent applications per year on a set of dummy variables that track the timing of CERN impact on innovation output, controlling for several covariates. Since we want to be sure that the change in firms' patenting activity post-dated the beginning of the collaboration with CERN, all specifications include also the leads of the dummy variable marking the beginning of the procurement relationship. More precisely, the variable denoted as $CERN_{i,t}^{(-2,-1)}$ indicates that firm *i* will receive the first order from CERN in at most a couple of years.

Both the Poisson and the Negative Binomial models reported in Table 2 are well suited for capturing the count nature of the dependent variable. Empirical evidence presented at the bottom of the table highlights that our data might be over-dispersed and therefore violate the assumptions underlying the Poisson regression model. The Poisson model implies that the variance of the number of patents in each period is equal to its expected value during the same time frame (i.e. equidispersion). The Negative Binomial distribution includes the Poisson as a special case and allows for both under- and over-dispersion. Over-dispersed variables have variance greater than the expected value.

**Figure 4. Patents per firm: relative years**

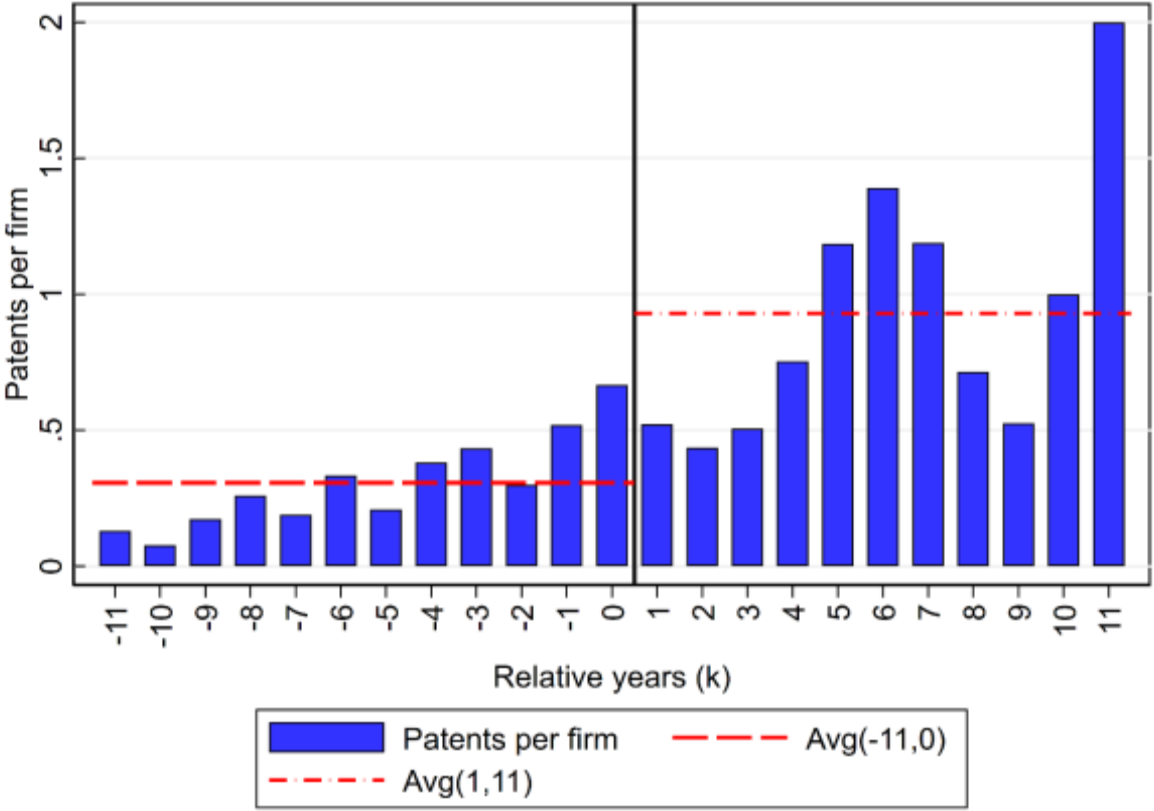

*Notes*: "Relative years (*k*)" measures the time distance from the first LHC order. Therefore, *k* > 0 indicates that the first order was received *k* years ago and *k* < 0 indicates that the order will be received in *k* years. For each *k* the figure reports the number of patents per firm. The dashed line denoted as "Avg(-11,0)" represents average number of patents per firm over relative years *k=-11,…,0*, that is before the start of the LHC procurement. Similarly, the dash-dotted line denoted as "Avg(1,11)" represents average number of patents per firm over relative years *k=1,…,11*, that is after the start of the LHC procurement.

Models in columns 1-3 and 4-6 of Table 2 include a growing number of control variables. The following comments refer to the most complete Negative Binomial specification in column 6. Focusing on the "CERN effect", we see that the coefficients on $CERN_{i,t}^k$ are statistically significant at the 95% confidence level only for $5 \leq k \leq 8$. This suggests once again that there is a considerable time lag separating LHC procurement from changes in number of patent applications. Moreover, the coefficient on $CERN_{i,t}^{(-2,-1)}$ is never statistically distinguishable from zero. This is consistent with the expectation that any impact of LHC procurement on firms' innovation output should post-date the start of the relationship with CERN and strengthen the causal interpretation of our results.

Focusing on control variables, we see that the coefficient on "*Hi-Tech$_i$*" is positive and statistically significant at the 95% confidence level, meaning that hi-tech firms on average file more patents than lo-tech ones. Size is also positively associated with the average number of patent applications. Similarly, we do observe a positive and statistically significant relationship between our proxy of R&D expenditure (i.e. "*IFA$_{it}$*") and firms' innovation output. This finding is in line with the vast literature on the R&D-patent nexus (Crépon et al. 1998; Gurmu and Pérez-Sebastiàn, 2008; Hausman et al., 1984). On the other hand, the remaining control variables (i.e. "*Avg. p$_i$*", "*Order$_i$*", "*pct$_{c,t}$*") are never statistically distinguishable from zero. Lastly, the model also includes a set of country, sector and time fixed effects. The null hypothesis that (each group of) these fixed effects are jointly equal to zero can always be rejected.

Comparison with the other columns of Table 2 shows that our main result - the existence of a "CERN effect" on firms' innovation output - is robust to both changes in the distributional assumptions (i.e. when considering the Poisson specifications) and to modifications of the set of control variables included in the model. The estimates and significance of the coefficients on the control variables are largely unaffected by to such changes.

All in all, Table 2 is consistent with the findings of Table 1, and further confirms *H1*. The "CERN effect" is associated with an increase in both the hazard to file a patent for the first time and in patent applications. In both cases, the effect shows up with a sizeable time lag, as envisaged by *H2*. A noticeable difference is that while the technological classification of firms leads to a statistically significant increase in the number of patent applications, the coefficient on this variable is statistically not distinguishable from zero in the Cox proportional hazard model.

**Table 2. Count data models**

|  | (1) Poisson | (2) Poisson | (3) Poisson | (4) NegBin | (5) NegBin | (6) NegBin |
|---|---|---|---|---|---|---|
| $CERN_{i,t}^{(-2,-1)}$ | 0.164 (0.214) | 0.167 (0.204) | 0.176 (0.207) | 0.122 (0.229) | 0.151 (0.220) | 0.159 (0.217) |
| $CERN_{i,t}^{0}$ | 0.637** (0.324) | 0.543 (0.334) | 0.542 (0.348) | 0.546* (0.315) | 0.564* (0.323) | 0.573* (0.315) |
| $CERN^{(1,2)}$ | 0.271 (0.236) | 0.145 (0.257) | 0.156 (0.278) | 0.191 (0.255) | 0.146 (0.253) | 0.158 (0.247) |
| $CERN^{(3,4)}$ | 0.482 (0.341) | 0.386 (0.352) | 0.410 (0.359) | 0.226 (0.311) | 0.177 (0.296) | 0.199 (0.276) |
| $CERN^{(5,6)}$ | 1.293*** (0.449) | 1.147** (0.487) | 1.209** (0.501) | 0.954** (0.419) | 0.927** (0.397) | 0.958** (0.394) |
| $CERN_{i,t}^{(7,8)}$ | 1.021** (0.452) | 0.934* (0.510) | 0.955 (0.612) | 1.132** (0.506) | 1.190** (0.498) | 1.213*** (0.471) |
| $CERN_{i,t}^{>8}$ | 0.808 (0.544) | 0.479 (0.634) | 0.506 (0.709) | 0.187 (0.508) | 0.066 (0.549) | 0.099 (0.607) |
| *Hi-Tech*$_i$ | 0.426** (0.193) | 0.653** (0.307) | 0.673** (0.308) | 0.632** (0.259) | 0.782*** (0.283) | 0.797** (0.317) |
| *Medium*$_i$ | 3.098** (1.549) | 2.982** (1.466) | 2.992** (1.449) | 3.301** (1.356) | 3.221*** (1.250) | 3.222*** (1.244) |
| *Large*$_i$ | 2.983** (1.492) | 2.821* (1.456) | 2.832* (1.467) | 2.921** (1.311) | 2.901** (1.225) | 2.909** (1.239) |
| *V Large*$_i$ | 4.787*** (1.665) | 4.446*** (1.647) | 4.461*** (1.677) | 4.561*** (1.488) | 4.279*** (1.448) | 4.288*** (1.472) |
| *Avg. p*$_i$ | -0.041 (0.559) | -0.017 (0.603) | -0.022 (0.614) | 0.037 (0.470) | 0.138 (0.512) | 0.132 (0.511) |
| *IFA*$_{it}$ |  | 0.105* (0.059) | 0.108* (0.059) |  | 0.113** (0.050) | 0.114** (0.046) |
| *Order*$_i$ |  |  | -0.011 (0.097) |  |  | -0.009 (0.105) |
| $pct_{c,t}$ |  |  | -0.030 (0.024) |  |  | -0.012 (0.036) |
| Overdispersion ($\delta$) |  |  |  | 2.1138 | 1.9905 | 1.9858 |
| $\alpha_c$ | 0.0000 | 0.0000 | 0.0000 | 0.0000 | 0.0000 | 0.0000 |
| $\alpha_s$ | 0.0000 | 0.0000 | 0.0000 | 0.0000 | 0.0000 | 0.0000 |
| $\alpha_t$ | 0.0000 | 0.0088 | 0.0000 | 0.0011 | 0.0000 | 0.0000 |

*Notes*: country ($\alpha_c$), sector ($\alpha_s$), year ($\alpha_t$) fixed effects have been included in all the specifications. * p-value < 0.10, ** p-value < 0.05, *** p-value < 0.01. Cluster robust standard errors (i.e. cluster is the 2 digits NACE code) in parentheses. The reference category are Belgian small-sized firms operating in the Other Manufacturing sector in 1993. The regression also includes a dummy variable equal to one when there is a missing value in IFA. "Overdispersion" denotes the estimate of the variance inflation factor in the Negative Binomial model where the variance is assumed to be equal to: var = $\delta \times$mean. The over dispersion test for H$_0$: $\delta =1$ is: 298.13 (0.0000), 274.01 (0.0000), 271.27 (0.0000), where p-values are shown in ().

*Robustness checks.* Baseline results in Table 2 rely on a sample design that excludes firms reporting more than 4 missing values in $IFA_{it}$. As a robustness check, we changed this arbitrary threshold from 4 up to 6 missing values per firm. Setting the exclusion rule to 5 missing observations yields a sample of 112 firms (or 1568 observations), while considering at most 6 missing values of the variable results in a sample of 121 firms (or 1694 observations). Our baseline results are robust to changes in the threshold used to exclude firms because of missing values in $IFA_{it}$. Similarly, excluding dichotomous variables acting as leads do not affect the estimated coefficients of neither the set of CERN dummies nor of control variables.

We also considered a completely different sample design in which we do not exclude firms that have never filed a patent. Adding entities without patents yields to a sample with 263 firms or 3682 observations. While carrying out the analysis including also companies that have never filed any patent leads to a much larger sample and possibly more efficient estimates, our main conclusions concerning the effect of CERN procurement on patents are not qualitatively different.

Lastly, we also entertained the same robustness checks considered in Section 4.1. First, we changed the proxy of firms' technological intensity from a dichotomous to a continuous variable. Second, we replaced the total amount of LHC orders with the (logarithm of the) count of such orders as an alternative proxy of the involvement and continuity of the procurement relationship with CERN. In both cases our main results remain unchanged. See Section A3.2 of the Appendix.

## 5. Conclusions

This paper contributes to the literature on the impact of BSC on technological innovation. A discussion of the earlier literature and qualitative evidence from case studies suggests that the main benefits for technological suppliers are learning and reputation, but there are also costs in terms of R&D and other investments with uncertain returns.

We have exploited a unique dataset to empirically test the impact of CERN procurement on the innovative output of its industrial partners. Patent applications have been used as a proxy of innovation output to investigate two related research questions. First, survival analysis has been implemented to assess whether the collaboration with CERN has increased the hazard of filing a patent for the first time. Second, count data models provide estimates of the impact of CERN procurement on the number of patents filed by CERN suppliers. Since collaborating firms have received their first order from CERN over a long time-span, we have a natural partition of statistical units into "suppliers" and "not-yet-suppliers" that allows to investigate the timing of the "CERN effect".

Our results show that a "CERN effect" does exist and is associated with an increase in both the hazard to file a patent for the first time and in patent applications. This effect is statistically significant only with a delay of some years (5-8) from the beginning of the procurement relationship. The existence of such a lag between procurement from BSC and innovation might signal that learning from technology at the frontier of science and translating such knowledge into commercial applications is a medium-long run process.

There is an important consideration in this perspective. As it has been highlighted by case studies, many suppliers facing entirely new technological problems need to initially rely on the expertise of people at CERN and to closely collaborate with them. This was the main result of Florio et al. (2018), who found that the impact on firms is higher the greater the relational governance of the procurement. Sometimes the R&D is implemented by the CERN in the first place and during the procurement relationship firms absorb radically new concepts and solutions required for the scientific purposes. After the end of the contract, firms reconsider what they have learnt and have to understand where new market opportunities may arise in the future. Only at the end of this "learning by interacting process", industrial partners' own R&D, and the subsequent commercial application of the new product/process developed, may start.

The crucial point is that this process is quite different from the one driving the established R&D-patents correlation in the usual business environment (such as in the seminal paper by Hausman et al. 1984) as it requires patient firms to invest to find a commercial application for the new knowledge arising from BSC procurement. What makes procurement for innovation at the frontier of science special is that it poses new technological challenges, triggering a sort of 'surprise' learning mechanism for firms, in the meaning of Solow (1997). This has potentially interesting implications as a complement to other, more established, innovation policies.

**Acknowledgments**

The authors thank participants at: the workshop on "The economic impact of CERN colliders: Technological spillovers from LHC to HL-LHC and beyond" at the CERN FCC Week held in Berlin in May 2017, at the XXX Conference of the Italian Society of Public Economics (SIEP) held in September 2018 at the University of Padua, and at the XIV Workshop of the Italian Society of Industrial Policy and Economics (SIEPI) held in January 2019 at the Roma Tre University . Moreover, we thank Emanuela Sirtori, who coordinated the interviews for the CERN&CSIL (2019) reports for useful comments on a previous version of the paper. Financial support from the FCC study in the frame of the Collaboration Agreement between the University of Milan and CERN (KE3044/ATS) is also gratefully acknowledged.

# Appendix

## A1. List of the companies involved in the 28 case-studies

| Company | Country | Size | Product | Application domain |
|---|---|---|---|---|
| 3M (SCHWEIZ) GMBH | CH | Large | Cables, insulating products, cooling liquids, safety solutions | Industry, safety, healthcare, electronics and energy, general consumption |
| A.SILVA MATOS - METALOMECANICA | PT | Large | Energy storage and production | Industry |
| AEMTEC GMBH | DE | Medium | Advanced microsystems and optoelectronics | Industry |
| ALIBAVA SYSTEMS SL. | ES | Small | Systems for radiation detectors | Science, Medicine, Industrial engineering |
| B & S INTERNATIONAL FRANCE | FR | Small | Development of electronic and electro-mechanical systems | Science |
| BAYARDS ALUMINIUM CONSTRUCTIE | NL | Large | Magnet and Nuclear technology | Science, Energy |
| BILFINGER NOELL GMBH (ex Babcock Noell) | DE | Large | Aluminium structures | Industry, civil engineering |
| BRUKER EAS GMBH | DE | Large | Advanced Superconductor Solutions | Research, industry, healthcare, energy |
| C.A.E.N. – SPA | IT | Large | Electronic instrumentation | Physics research, material science industry, medical industry, homeland security |
| CARBOLITE GERO GMBH & CO. KG | DE | Small | High-tech cryogenic systems | Science, industry, aerospace |
| CRIOTEC IMPIANTI S.P.A. | IT | Small | Flywheels and scientific installation components | Science, railway industry, energy |
| ELYTT ENERGY | ES | Small | Supervisory Control and Data Acquisition Software | - |
| ETM Professional Controls | AT | Large | Civil engineering for complex projects | Civil engineering |
| GEOCONSULT ZT GMBH | AT | Large | High-temperature furnaces and ovens manufacturer | Industry, science |
| HAMAMATSU PHOTONICS KK | ES | Large | Photonics Technology | Industry, science, healthcare |
| HEINZINGER ELECTRONIC GMBH | DE | Small | Precision high-current and high-voltage power supplies | Industry, science |
| LEYBOLD SCHWEIZ AG | CH | Large | Vacuum components and systems | Science, industry, food & beverage |
| LINDE KRYOTECHNIK AG | CH | Small | Helium and hydrogen liquefiers and refrigerators | Industry, science, healthcare |
| M & I MATERIALS LIMITED | GB | Medium | Commercialising Materials for Demanding Applications | Science, aerospace, healthcare |
| NATIONAL INSTRUMENTS SWITZERLAND | CH | Large | Software and Hardware Platforms | Aerospace and Defence, Energy, Automotive, Industry machinery, Wireless Communications, Science |
| NOVAPACK TECHNOLOGIES SARL | FR | Small | Packaging solutions for electronic components | Science, Industry |
| OCEM POWER ELECTRONICS | IT | Medium | Power suppliers and converters | science, industry, healthcare, transport |
| OPTIM WAFER SERVICES | FR | Small | Wafer processing services | - |
| PFEIFFER VACUUM TECHNOLOGY AG | DE | Large | Vacuum generation, measurement and analysis | Industry, science |
| SERTEC | ES | Small | Engineering services, test and instrumentation systems | Science, aerospace, railway |
| SIGMAPHI | FR | Large | Particle accelerators components and superconducting systems | Science, industry, healthcare |
| SIMIC S.P.A. | IT | Large | Cryostats and vacuum vessels | Industry, science, oil & gas, energy |
| VOEST-ALPINE AG | AT | Large | High quality steel for sophisticated applications | Automotive, aerospace, oil and gas industry, consumer good industry |

**A2. Classification of hi- and lo-tech firms**

In order to assign suppliers to the high-tech or low-tech group, we took advantage of the fact that in the original database CERN orders are classified by an "activity code" identifying each product type with a highly detailed 3-digit level. We used the 2-digit classification, which covers around 100 items and was sufficiently detailed for our purposes. In some cases, we also inspected the 3-digit classification to better interpret the technological content.

After a preliminary analysis of the overall distribution of order codes, we followed Florio et al. (2016) in identifying the specific activity codes most likely to be associated with high-tech goods and services for the construction of the LHC. In some instances, the code descriptors were generic ("28-Electrical engineering," say, or "45-Software"). To minimise classification errors, we sampled 300 orders for a more in-depth analysis. These orders were placed with 207 different suppliers, 16% of all those who received at least one order for the LHC during the period under analysis. The orders thus sampled were then evaluated in detail by CERN experts and classified, according to their technological intensity, along a five-point scale designed to capture differences in both product specificity and closeness of the supplier's collaboration with CERN:

- Class 1: most likely "off-the-shelf" orders of low technological intensity;
- Class 2: off-the-shelf orders with average technological intensity;
- Class 3: mostly off-the-shelf but usually high-tech and requiring some careful specification;
- Class 4: high-tech orders with moderate to high intensity of specification activity to customise products for the LHC;
- Class 5: products at the technological frontier, with intensive customisation and co-design involving CERN staff.

We defined high-tech codes as Classes 3, 4 and 5 and then divided the LHC suppliers into two broad groups, according to their opportunity to deliver high-tech orders in the initial procurement event. According to the activity code assigned to the first order, 68.5% of our sample companies are part of the high-tech category, with a slight over-representation of 7.5 percentage points in relation to the original CERN data (61%). There is some risk of misclassification, in that non-high-tech companies may have gained the ability, over time, to satisfy high-tech orders, and that many companies received more than one order, which are not necessarily all coded alike. However, the data indicate that the first order is generally a good predictor of the technological intensity of subsequent ones.

# A3. Robustness checks

## A3.1 Survival analysis: further results

**Table A1. Cox proportional hazard model: alternative hi-tech classification**

|  | (1) $k=0$ | (2) $k=1$ | (3) $k=2$ | (4) $k=3$ | (5) $k=4$ | (6) $k=5$ | (7) $k=6$ | (8) $k=7$ | (9) $k=8$ | (10) $k \geq 9$ |
|---|---|---|---|---|---|---|---|---|---|---|
| $CERN_{i,t}^k$ | 0.034 | 0.044* | 0.047** | 0.054** | 0.068*** | 0.081*** | 0.089*** | 0.072** | -0.009 | 0.034 |
|  | (0.028) | (0.026) | (0.024) | (0.023) | (0.024) | (0.025) | (0.029) | (0.033) | (0.047) | (0.052) |
| $\%HTO_i$ | 0.179 | 0.181 | 0.185 | 0.184 | 0.188 | 0.195 | 0.201 | 0.196 | 0.177 | 0.180 |
|  | (0.222) | (0.222) | (0.222) | (0.222) | (0.222) | (0.222) | (0.222) | (0.222) | (0.222) | (0.223) |
| $Order_i$ | 0.003 | -0.001 | -0.002 | -0.004 | -0.008 | -0.011 | -0.007 | 0.002 | 0.014 | 0.012 |
|  | (0.046) | (0.046) | (0.046) | (0.046) | (0.046) | (0.046) | (0.046) | (0.046) | (0.046) | (0.045) |
| $Medium_i$ | 0.515 | 0.517 | 0.518 | 0.523 | 0.530 | 0.534 | 0.530 | 0.523 | 0.520 | 0.520 |
|  | (0.347) | (0.346) | (0.346) | (0.346) | (0.346) | (0.346) | (0.347) | (0.348) | (0.350) | (0.350) |
| $Large_i$ | 1.162*** | 1.160*** | 1.159*** | 1.164*** | 1.171*** | 1.175*** | 1.167*** | 1.162*** | 1.171*** | 1.167*** |
|  | (0.343) | (0.342) | (0.343) | (0.343) | (0.343) | (0.344) | (0.344) | (0.346) | (0.346) | (0.346) |
| $V\,Large_i$ | 2.053*** | 2.055*** | 2.060*** | 2.070*** | 2.089*** | 2.101*** | 2.095*** | 2.071*** | 2.049*** | 2.050*** |
|  | (0.367) | (0.365) | (0.365) | (0.365) | (0.366) | (0.366) | (0.367) | (0.367) | (0.369) | (0.369) |
| $pct_{c,t}$ | 0.002 | 0.002 | 0.001 | 0.001 | 0.002 | 0.001 | 0.002 | 0.002 | 0.002 | 0.002 |
|  | (0.003) | (0.003) | (0.003) | (0.003) | (0.003) | (0.003) | (0.003) | (0.003) | (0.003) | (0.003) |
| $\alpha_c$ | yes | yes | yes | yes | yes | yes | yes | yes | yes | yes |
| $\alpha_s$ | yes | yes | yes | yes | yes | yes | yes | yes | yes | yes |

\* p-value < 0.10, ** p-value < 0.05, *** p-value < 0.01

Notes: country ($\alpha_c$) and sector ($\alpha_s$) fixed effects have been included in all the specifications.. Heteroskedastic robust standard errors in parentheses. %$HTO_i$ denotes the share of hi-tech orders on the total number of orders received.

**Table A2. Cox proportional hazard model: no. orders in place of total order amount**

|  | (1) $k=0$ | (2) $k=1$ | (3) $k=2$ | (4) $k=3$ | (5) $k=4$ | (6) $k=5$ | (7) $k=6$ | (8) $k=7$ | (9) $k=8$ | (10) $k \geq 9$ |
|---|---|---|---|---|---|---|---|---|---|---|
| $CERN_{i,t}^k$ | 0.039 | 0.044* | 0.047** | 0.054** | 0.068*** | 0.081*** | 0.089*** | 0.072** | -0.009 | 0.034 |
|  | (0.028) | (0.026) | (0.024) | (0.023) | (0.024) | (0.025) | (0.029) | (0.033) | (0.047) | (0.052) |
| $Hi\text{-}tech_i$ | 0.144 | 0.181 | 0.185 | 0.184 | 0.188 | 0.195 | 0.201 | 0.196 | 0.177 | 0.180 |
|  | (0.215) | (0.222) | (0.222) | (0.222) | (0.222) | (0.222) | (0.222) | (0.222) | (0.222) | (0.223) |
| $\#Order_i$ | -0.073 | -0.001 | -0.002 | -0.004 | -0.008 | -0.011 | -0.007 | 0.002 | 0.014 | 0.012 |
|  | (0.080) | (0.046) | (0.046) | (0.046) | (0.046) | (0.046) | (0.046) | (0.046) | (0.046) | (0.045) |
| $Medium_i$ | 0.464 | 0.517 | 0.518 | 0.523 | 0.530 | 0.534 | 0.530 | 0.523 | 0.520 | 0.520 |
|  | (0.340) | (0.346) | (0.346) | (0.346) | (0.346) | (0.346) | (0.347) | (0.348) | (0.350) | (0.350) |
| $Large_i$ | 1.151*** | 1.160*** | 1.159*** | 1.164*** | 1.171*** | 1.175*** | 1.167*** | 1.162*** | 1.171*** | 1.167*** |
|  | (0.338) | (0.342) | (0.343) | (0.343) | (0.343) | (0.344) | (0.344) | (0.346) | (0.346) | (0.346) |
| $V\,Large_i$ | 2.041*** | 2.055*** | 2.060*** | 2.070*** | 2.089*** | 2.101*** | 2.095*** | 2.071*** | 2.049*** | 2.050*** |
|  | (0.355) | (0.365) | (0.365) | (0.365) | (0.366) | (0.366) | (0.367) | (0.367) | (0.369) | (0.369) |
| $pct_{c,t}$ | 0.002 | 0.002 | 0.001 | 0.001 | 0.002 | 0.001 | 0.002 | 0.002 | 0.002 | 0.002 |
|  | (0.003) | (0.003) | (0.003) | (0.003) | (0.003) | (0.003) | (0.003) | (0.003) | (0.003) | (0.003) |
| $\alpha_c$ | yes | yes | yes | yes | yes | yes | yes | yes | yes | yes |
| $\alpha_s$ | yes | yes | yes | yes | yes | yes | yes | yes | yes | yes |

\* p-value < 0.10, ** p-value < 0.05, *** p-value < 0.01

Notes: country ($\alpha_c$) and sector ($\alpha_s$) fixed effects have been included in all the specifications.. Heteroskedastic robust standard errors in parentheses. $\#Order_i$ denotes logarithm of the total number of LHC-related orders received by firm $i$.

## A3.2 Count data models: further results

**Table A3. Count data models – full sample**

|  | (1) Poisson | (2) Poisson | (3) Poisson | (4) NegBin | (5) NegBin | (6) NegBin |
|---|---|---|---|---|---|---|
| $CERN_{i,t}^{(-2,-1)}$ | 0.259 (0.167) | 0.247 (0.161) | 0.229 (0.180) | 0.126 (0.207) | 0.154 (0.193) | 0.130 (0.189) |
| $CERN_{i,t}^{0}$ | 0.766*** (0.217) | 0.641*** (0.222) | 0.603** (0.249) | 0.522** (0.250) | 0.516** (0.257) | 0.483** (0.216) |
| $CERN_{i,t}^{(1,2)}$ | 0.442*** (0.150) | 0.292* (0.155) | 0.251 (0.212) | 0.247 (0.297) | 0.178 (0.283) | 0.135 (0.270) |
| $CERN_{i,t}^{(3,4)}$ | 0.687** (0.268) | 0.593** (0.274) | 0.549* (0.296) | 0.355 (0.326) | 0.298 (0.304) | 0.243 (0.327) |
| $CERN^{(5,6)}$ | 1.528*** (0.284) | 1.394*** (0.292) | 1.368*** (0.357) | 1.012*** (0.348) | 0.963*** (0.304) | 0.899*** (0.316) |
| $CERN_{i,t}^{(7,8)}$ | 1.297*** (0.348) | 1.264*** (0.376) | 1.165** (0.553) | 1.352*** (0.404) | 1.410*** (0.419) | 1.303*** (0.390) |
| $CERN_{i,t}^{>8}$ | 1.076*** (0.337) | 0.807** (0.355) | 0.721 (0.516) | 0.235 (0.574) | 0.177 (0.596) | 0.083 (0.732) |
| $Hi\text{-}Tech_i$ | 0.534* (0.284) | 0.690** (0.284) | 0.648*** (0.247) | 0.641* (0.336) | 0.796** (0.319) | 0.763** (0.329) |
| $Medium_i$ | 2.659** (1.281) | 2.596** (1.216) | 2.631** (1.172) | 2.707** (1.233) | 2.651** (1.155) | 2.671** (1.117) |
| $Large_i$ | 3.005** (1.256) | 2.926** (1.267) | 2.932** (1.248) | 3.082*** (1.144) | 3.025*** (1.137) | 3.033*** (1.130) |
| $V\ Large_i$ | 5.519*** (1.395) | 5.173*** (1.425) | 5.123*** (1.477) | 5.274*** (1.254) | 4.890*** (1.286) | 4.855*** (1.321) |
| $Avg.\ p_i$ | 0.184 (0.430) | 0.250 (0.438) | 0.286 (0.411) | 0.383 (0.411) | 0.516 (0.427) | 0.536 (0.399) |
| $IFA_{it}$ |  | 0.121** (0.049) | 0.121** (0.048) |  | 0.133*** (0.051) | 0.133*** (0.052) |
| $Order_i$ |  |  | 0.035 (0.118) |  |  | 0.034 (0.124) |
| $pct_{c,t}$ |  |  | -0.028 (0.024) |  |  | -0.020 (0.034) |
| Overdispersion ($\delta$) |  |  |  | 3.5123 | 3.3061 | 3.3070 |
| $\alpha_c$ | 0.0000 | 0.0000 | 0.0000 | 0.0000 | 0.0000 | 0.0000 |
| $\alpha_s$ | 0.0000 | 0.0000 | 0.0000 | 0.0000 | 0.0000 | 0.0000 |
| $\alpha_t$ | 0.0004 | 0.0000 | 0.0000 | 0.0000 | 0.0000 | 0.0000 |

\* p<0.1, ** p<0.05, *** p<0.01.
Notes: Std. Err. clustered by sector (i.e. 2 digits NACE code) in parentheses; "Country FE" ("Sector FE" / "Year FE") shows the p-value associated with the Wald test of the joint null hypothesis that country (sector / year) fixed effects are not statistically significant.
Overdispersion tests (p-values): 414.90 (0.0000), 380.55 (0.0000), 377.23 (0.0000). The regression also includes a dummy variable equal to one when there is a missing value in IFA.

**Table A4. Count data models - without leads**

|  | (1) Poisson | (2) Poisson | (3) Poisson | (4) NegBin | (5) NegBin | (6) NegBin |
|---|---|---|---|---|---|---|
| $CERN_{i,t}^{0}$ | 0.555** | 0.459* | 0.452* | 0.481* | 0.484* | 0.487* |
|  | (0.253) | (0.259) | (0.268) | (0.255) | (0.264) | (0.258) |
| $CERN_{i,t}^{(1,2)}$ | 0.179 | 0.051 | 0.053 | 0.122 | 0.060 | 0.066 |
|  | (0.167) | (0.178) | (0.195) | (0.186) | (0.185) | (0.180) |
| $CERN_{i,t}^{(3,4)}$ | 0.378 | 0.279 | 0.293 | 0.142 | 0.074 | 0.087 |
|  | (0.271) | (0.262) | (0.262) | (0.232) | (0.200) | (0.179) |
| $CERN^{(5,6)}$ | 1.181*** | 1.031*** | 1.082*** | 0.863*** | 0.816*** | 0.835*** |
|  | (0.372) | (0.393) | (0.398) | (0.330) | (0.293) | (0.291) |
| $CERN_{i,t}^{(7,8)}$ | 0.901** | 0.812* | 0.819 | 1.036** | 1.071*** | 1.081*** |
|  | (0.372) | (0.433) | (0.523) | (0.432) | (0.413) | (0.380) |
| $CERN_{i,t}^{>8}$ | 0.688 | 0.356 | 0.369 | 0.090 | -0.054 | -0.034 |
|  | (0.448) | (0.529) | (0.593) | (0.436) | (0.472) | (0.524) |
| $Hi\text{-}Tech_i$ | 0.433** | 0.660** | 0.675** | 0.639** | 0.788*** | 0.799** |
|  | (0.189) | (0.303) | (0.307) | (0.256) | (0.281) | (0.317) |
| $Medium_i$ | 3.059** | 2.942** | 2.954** | 3.272** | 3.186** | 3.187*** |
|  | (1.533) | (1.450) | (1.429) | (1.337) | (1.238) | (1.223) |
| $Large_i$ | 2.954** | 2.791* | 2.800* | 2.899** | 2.875** | 2.881** |
|  | (1.475) | (1.438) | (1.442) | (1.293) | (1.213) | (1.219) |
| $V\,Large_i$ | 4.757*** | 4.414*** | 4.424*** | 4.543*** | 4.260*** | 4.265*** |
|  | (1.652) | (1.630) | (1.651) | (1.481) | (1.446) | (1.463) |
| $Avg.\ p_i$ | -0.054 | -0.030 | -0.032 | 0.031 | 0.129 | 0.126 |
|  | (0.545) | (0.588) | (0.598) | (0.469) | (0.511) | (0.508) |
| $IFA_{it}$ |  | 0.106* | 0.109* |  | 0.112** | 0.112** |
|  |  | (0.059) | (0.058) |  | (0.051) | (0.048) |
| $Order_i$ |  |  | -0.007 |  |  | -0.006 |
|  |  |  | (0.097) |  |  | (0.104) |
| $pct_{c,t}$ |  |  | -0.030 |  |  | -0.012 |
|  |  |  | (0.024) |  |  | (0.035) |
| Overdispersion ($\delta$) |  |  |  | 2.1146 | 1.9917 | 1.9878 |
| $\alpha_c$ | 0.0000 | 0.0000 | 0.0000 | 0.0000 | 0.0000 | 0.0000 |
| $\alpha_s$ | 0.0000 | 0.0000 | 0.0000 | 0.0000 | 0.0000 | 0.0000 |
| $\alpha_t$ | 0.0000 | 0.0046 | 0.0000 | 0.0000 | 0.0000 | 0.0000 |

\* p<0.1, ** p<0.05, *** p<0.01 Notes: Std. Err. clustered by sector (i.e. 2 digits NACE code) in parentheses; "Country FE" ("Sector FE" / "Year FE") shows the p-value associated with the Wald test of the joint null hypothesis that country (sector / year) fixed effects are not statistically significant. Overdispersion tests (p-values): 299.04 (0.0000), 274.81 (0.0000), 272.15 (0.0000). The regression also includes a dummy variable equal to one when there is a missing value in IFA.

**Table A5. Count data models - larger sample (5 missing values in the $IFA_{it}$)**

|  | (1) Poisson | (2) Poisson | (3) Poisson | (4) NegBin | (5) NegBin | (6) NegBin |
|---|---|---|---|---|---|---|
| $CERN_{i,t}^{(-2,-1)}$ | 0.249 | 0.246 | 0.241 | 0.207 | 0.231 | 0.227 |
|  | (0.201) | (0.197) | (0.202) | (0.214) | (0.213) | (0.213) |
| $CERN_{i,t}^{0}$ | 0.699** | 0.596* | 0.583* | 0.596* | 0.600* | 0.595* |
|  | (0.325) | (0.333) | (0.347) | (0.339) | (0.346) | (0.341) |
| $CERN_{i,t}^{(1,2)}$ | 0.271 | 0.142 | 0.134 | 0.212 | 0.165 | 0.157 |
|  | (0.241) | (0.261) | (0.290) | (0.271) | (0.269) | (0.273) |
| $CERN_{i,t}^{(3,4)}$ | 0.509 | 0.396 | 0.394 | 0.272 | 0.214 | 0.200 |
|  | (0.354) | (0.369) | (0.381) | (0.339) | (0.330) | (0.327) |
| $CERN^{(5,6)}$ | 1.374*** | 1.215** | 1.237** | 1.031** | 1.002** | 0.982** |
|  | (0.450) | (0.484) | (0.505) | (0.444) | (0.424) | (0.434) |
| $CERN_{i,t}^{(7,8)}$ | 1.044** | 0.940* | 0.921 | 1.111* | 1.156** | 1.142** |
|  | (0.488) | (0.535) | (0.641) | (0.588) | (0.572) | (0.565) |
| $CERN_{i,t}^{>8}$ | 0.846 | 0.532 | 0.520 | 0.263 | 0.126 | 0.100 |
|  | (0.518) | (0.580) | (0.660) | (0.472) | (0.496) | (0.592) |
| $Hi\text{-}Tech_i$ | 0.471** | 0.699** | 0.694** | 0.642** | 0.790*** | 0.780** |
|  | (0.191) | (0.294) | (0.293) | (0.270) | (0.289) | (0.329) |
| $Medium_i$ | 3.299** | 3.184** | 3.200** | 3.489*** | 3.427*** | 3.424*** |
|  | (1.481) | (1.393) | (1.370) | (1.297) | (1.191) | (1.176) |
| $Large_i$ | 2.940** | 2.779** | 2.785** | 2.909** | 2.888** | 2.882** |
|  | (1.467) | (1.418) | (1.411) | (1.286) | (1.201) | (1.203) |
| $V\,Large_i$ | 4.822*** | 4.482*** | 4.480*** | 4.572*** | 4.314*** | 4.309*** |
|  | (1.637) | (1.614) | (1.634) | (1.474) | (1.440) | (1.449) |
| $Avg.\,p_i$ | -0.063 | -0.038 | -0.026 | 0.086 | 0.182 | 0.185 |
|  | (0.492) | (0.511) | (0.503) | (0.434) | (0.471) | (0.459) |
| $IFA_{it}$ |  | 0.103* | 0.104** |  | 0.103** | 0.102** |
|  |  | (0.055) | (0.053) |  | (0.047) | (0.043) |
| $Order_i$ |  |  | 0.006 |  |  | 0.006 |
|  |  |  | (0.100) |  |  | (0.108) |
| $pct_{c,t}$ |  |  | -0.023 |  |  | 0.008 |
|  |  |  | (0.027) |  |  | (0.033) |
| Overdispersion ($\delta$) |  |  |  | 2.2129 | 2.0988 | 2.1012 |
| $\alpha_c$ | 0.0000 | 0.0000 | 0.0000 | 0.0000 | 0.0000 | 0.0000 |
| $\alpha_s$ | 0.0000 | 0.0000 | 0.0000 | 0.0000 | 0.0000 | 0.0000 |
| $\alpha_t$ | 0.0074 | 0.0892 | 0.0003 | 0.0066 | 0.0000 | 0.0000 |

\* p<0.1, \*\* p<0.05, \*\*\* p<0.01
Notes: Std. Err. clustered by sector (i.e. 2 digits NACE code) in parentheses; "Country FE" ("Sector FE" / "Year FE") shows the p-value associated with the Wald test of the joint null hypothesis that country (sector / year) fixed effects are not statistically significant.
Overdispersion tests (p-values): 320.39 (0.0000), 295.94 (0.0000), 294.26 (0.0000). The regression also includes a dummy variable equal to one when there is a missing value in IFA.

**Table A6. Count data models - larger sample (6 missing values in the $IFA_{it}$)**

|  | (1) Poisson | (2) Poisson | (3) Poisson | (4) NegBin | (5) NegBin | (6) NegBin |
|---|---|---|---|---|---|---|
| $CERN_{i,t}^{(-2,-1)}$ | 0.287* | 0.299* | 0.286* | 0.211 | 0.248 | 0.233 |
|  | (0.173) | (0.167) | (0.169) | (0.201) | (0.198) | (0.190) |
| $CERN_{i,t}^{0}$ | 0.786*** | 0.714*** | 0.691** | 0.660** | 0.683** | 0.663** |
|  | (0.261) | (0.249) | (0.270) | (0.268) | (0.276) | (0.261) |
| $CERN_{i,t}^{(1,2)}$ | 0.421** | 0.318 | 0.295 | 0.337 | 0.304 | 0.279 |
|  | (0.195) | (0.207) | (0.223) | (0.251) | (0.251) | (0.237) |
| $CERN_{i,t}^{(3,4)}$ | 0.649* | 0.573* | 0.553* | 0.362 | 0.329 | 0.292 |
|  | (0.347) | (0.348) | (0.319) | (0.331) | (0.324) | (0.276) |
| $CERN^{(5,6)}$ | 1.524*** | 1.404*** | 1.410*** | 1.100** | 1.087*** | 1.044*** |
|  | (0.417) | (0.426) | (0.400) | (0.431) | (0.403) | (0.358) |
| $CERN_{i,t}^{(7,8)}$ | 1.192*** | 1.143*** | 1.099** | 1.173** | 1.260** | 1.212*** |
|  | (0.359) | (0.381) | (0.486) | (0.507) | (0.515) | (0.461) |
| $CERN_{i,t}^{>8}$ | 0.999** | 0.760 | 0.723 | 0.346 | 0.268 | 0.208 |
|  | (0.494) | (0.509) | (0.553) | (0.501) | (0.525) | (0.556) |
| $Hi\text{-}Tech_i$ | 0.576*** | 0.775*** | 0.762*** | 0.767*** | 0.900*** | 0.881*** |
|  | (0.191) | (0.289) | (0.282) | (0.220) | (0.256) | (0.275) |
| $Medium_i$ | 3.567** | 3.464** | 3.470** | 3.660*** | 3.595*** | 3.594*** |
|  | (1.480) | (1.353) | (1.357) | (1.255) | (1.114) | (1.096) |
| $Large_i$ | 3.082** | 2.932** | 2.931** | 3.028** | 3.006*** | 3.001*** |
|  | (1.455) | (1.373) | (1.374) | (1.236) | (1.114) | (1.105) |
| $V\,Large_i$ | 4.950*** | 4.570*** | 4.558*** | 4.621*** | 4.318*** | 4.305*** |
|  | (1.624) | (1.594) | (1.613) | (1.420) | (1.367) | (1.359) |
| $Avg.\,p_i$ | 0.478 | 0.471 | 0.484 | 0.479 | 0.545 | 0.555 |
|  | (0.412) | (0.423) | (0.411) | (0.432) | (0.434) | (0.421) |
| $IFA_{it}$ |  | 0.105* | 0.105** |  | 0.113** | 0.111** |
|  |  | (0.054) | (0.053) |  | (0.051) | (0.049) |
| $Order_i$ |  |  | 0.013 |  |  | 0.019 |
|  |  |  | (0.084) |  |  | (0.087) |
| $pct_{c,t}$ |  |  | -0.023 |  |  | 0.002 |
|  |  |  | (0.028) |  |  | (0.036) |
| Overdispersion ($\delta$) |  |  |  | 2.2623 | 2.1493 | 2.1521 |
| $\alpha_c$ | 0.0000 | 0.0000 | 0.0000 | 0.0000 | 0.0000 | 0.0000 |
| $\alpha_s$ | 0.0000 | 0.0000 | 0.0000 | 0.0000 | 0.0000 | 0.0000 |
| $\alpha_t$ | 0.0145 | 0.0882 | 0.0317 | 0.0002 | 0.0000 | 0.0000 |

* p<0.1, ** p<0.05, *** p<0.01
Notes: Std. Err. clustered by sector (i.e. 2 digits NACE code) in parentheses; "Country FE" ("Sector FE" / "Year FE") shows the p-value associated with the Wald test of the joint null hypothesis that country (sector / year) fixed effects are not statistically significant. Overdispersion tests (p-values): 356.00 (0.0000), 332.26 (0.0000), 330.38 (0.0000). The regression also includes a dummy variable equal to one when there is a missing value in IFA.

**Table A7. Count data models - with no. of orders in place of order amount**

|  | (1) Poisson | (2) Poisson | (3) Poisson | (4) NegBin | (5) NegBin | (6) NegBin |
|---|---|---|---|---|---|---|
| $CERN_{i,t}^{(-2,-1)}$ | 0.164 (0.214) | 0.167 (0.204) | 0.094 (0.226) | 0.122 (0.229) | 0.151 (0.220) | 0.089 (0.231) |
| $CERN_{i,t}^{0}$ | 0.637** (0.324) | 0.543 (0.334) | 0.439 (0.363) | 0.546* (0.315) | 0.564* (0.323) | 0.478 (0.303) |
| $CERN_{i,t}^{(1,2)}$ | 0.271 (0.236) | 0.145 (0.257) | 0.020 (0.325) | 0.191 (0.255) | 0.146 (0.253) | 0.045 (0.270) |
| $CERN_{i,t}^{(3,4)}$ | 0.482 (0.341) | 0.386 (0.352) | 0.227 (0.393) | 0.226 (0.311) | 0.177 (0.296) | 0.032 (0.303) |
| $CERN^{(5,6)}$ | 1.293*** (0.449) | 1.147** (0.487) | 0.980* (0.547) | 0.954** (0.419) | 0.927** (0.397) | 0.747* (0.412) |
| $CERN_{i,t}^{(7,8)}$ | 1.021** (0.452) | 0.934* (0.510) | 0.651 (0.693) | 1.132** (0.506) | 1.190** (0.498) | 0.940** (0.474) |
| $CERN_{i,t}^{>8}$ | 0.808 (0.544) | 0.479 (0.634) | 0.195 (0.819) | 0.187 (0.508) | 0.066 (0.549) | -0.174 (0.728) |
| $Hi\text{-}Tech_i$ | 0.426** (0.193) | 0.653** (0.307) | 0.588** (0.249) | 0.632** (0.259) | 0.782*** (0.283) | 0.728** (0.286) |
| $Medium_i$ | 3.098** (1.549) | 2.982** (1.466) | 3.075** (1.298) | 3.301** (1.356) | 3.221*** (1.250) | 3.274*** (1.134) |
| $Large_i$ | 2.983** (1.492) | 2.821* (1.456) | 2.819** (1.348) | 2.921** (1.311) | 2.901** (1.225) | 2.904** (1.160) |
| $V\,Large_i$ | 4.787*** (1.665) | 4.446*** (1.647) | 4.427*** (1.525) | 4.561*** (1.488) | 4.279*** (1.448) | 4.289*** (1.350) |
| $Avg.\,p_i$ | -0.041 (0.559) | -0.017 (0.603) | 0.094 (0.500) | 0.037 (0.470) | 0.138 (0.512) | 0.197 (0.451) |
| $IFA_{it}$ |  | 0.105* (0.059) | 0.099* (0.053) |  | 0.113** (0.050) | 0.105** (0.044) |
| $\#Order_i$ |  |  | 0.156 (0.221) |  |  | 0.132 (0.219) |
| $pct_{c,t}$ |  |  | -0.029 (0.024) |  |  | -0.012 (0.036) |
| Overdispersion ($\delta$) |  |  |  | 2.1138 | 1.9905 | 1.9684 |
| $\alpha_c$ | 0.0000 | 0.0000 | 0.0000 | 0.0000 | 0.0000 | 0.0000 |
| $\alpha_s$ | 0.0000 | 0.0000 | 0.0000 | 0.0000 | 0.0000 | 0.0000 |
| $\alpha_t$ | 0.0000 | 0.0088 | 0.0000 | 0.0011 | 0.0000 | 0.0000 |

\* p<0.1, ** p<0.05, *** p<0.01

Notes: Std. Err. clustered by sector (i.e. 2 digits NACE code) in parentheses; "Country FE" ("Sector FE" / "Year FE") shows the p-value associated with the Wald test of the joint null hypothesis that country (sector / year) fixed effects are not statistically significant. Overdispersion tests (p-values): 298.13 (0.0000), 274.01 (0.0000), 265.79 (0.0000). The regression also includes a dummy variable equal to one when there is a missing value in IFA.

**Table A8. Count data models - alternative hi-tech classification**

|  | (1) Poisson | (2) Poisson | (3) Poisson | (4) NegBin | (5) NegBin | (6) NegBin |
|---|---|---|---|---|---|---|
| $CERN_{i,t}^{(-2,-1)}$ | 0.188 | 0.208 | 0.217 | 0.147 | 0.183 | 0.184 |
|  | (0.218) | (0.210) | (0.219) | (0.231) | (0.225) | (0.226) |
| $CERN_{i,t}^{0}$ | 0.666** | 0.593* | 0.590* | 0.578* | 0.605* | 0.606* |
|  | (0.327) | (0.332) | (0.353) | (0.320) | (0.328) | (0.323) |
| $CERN_{i,t}^{(1,2)}$ | 0.310 | 0.209 | 0.219 | 0.239 | 0.202 | 0.205 |
|  | (0.241) | (0.254) | (0.288) | (0.257) | (0.252) | (0.256) |
| $CERN_{i,t}^{(3,4)}$ | 0.533 | 0.471 | 0.494 | 0.288 | 0.252 | 0.259 |
|  | (0.348) | (0.355) | (0.379) | (0.322) | (0.302) | (0.297) |
| $CERN^{(5,6)}$ | 1.351*** | 1.241** | 1.306** | 1.031** | 1.021** | 1.032** |
|  | (0.452) | (0.483) | (0.518) | (0.433) | (0.410) | (0.420) |
| $CERN_{i,t}^{(7,8)}$ | 1.084** | 1.037** | 1.055* | 1.219** | 1.295*** | 1.293*** |
|  | (0.439) | (0.494) | (0.623) | (0.513) | (0.496) | (0.482) |
| $CERN_{i,t}^{>8}$ | 0.871* | 0.557 | 0.584 | 0.294 | 0.192 | 0.199 |
|  | (0.525) | (0.585) | (0.688) | (0.509) | (0.543) | (0.627) |
| $\%HTO_i$ | 0.510*** | 0.798*** | 0.816*** | 0.659** | 0.833** | 0.837** |
|  | (0.135) | (0.218) | (0.239) | (0.309) | (0.333) | (0.369) |
| $Medium_i$ | 3.209** | 3.153** | 3.170** | 3.373** | 3.315** | 3.320** |
|  | (1.628) | (1.532) | (1.515) | (1.428) | (1.326) | (1.312) |
| $Large_i$ | 3.086** | 2.977** | 2.992** | 2.992** | 2.991** | 2.997** |
|  | (1.538) | (1.482) | (1.504) | (1.358) | (1.277) | (1.290) |
| $V Large_i$ | 4.869*** | 4.558*** | 4.575*** | 4.613*** | 4.341*** | 4.344*** |
|  | (1.709) | (1.674) | (1.713) | (1.522) | (1.483) | (1.501) |
| $Avg. p_i$ | -0.020 | 0.017 | 0.015 | 0.041 | 0.148 | 0.149 |
|  | (0.591) | (0.649) | (0.653) | (0.489) | (0.538) | (0.526) |
| $IFA_{it}$ |  | 0.112** | 0.115** |  | 0.115** | 0.116*** |
|  |  | (0.053) | (0.052) |  | (0.046) | (0.043) |
| $Order_i$ |  |  | -0.009 |  |  | 0.000 |
|  |  |  | (0.098) |  |  | (0.104) |
| $pct_{c,t}$ |  |  | -0.032 |  |  | -0.013 |
|  |  |  | (0.025) |  |  | (0.035) |
| Overdispersion ($\delta$) |  |  |  | 2.0912 | 1.9560 | 1.9530 |
| $\alpha_c$ | 0.0000 | 0.0000 | 0.0000 | 0.0000 | 0.0000 | 0.0000 |
| $\alpha_s$ | 0.0000 | 0.0000 | 0.0000 | 0.0000 | 0.0000 | 0.0000 |
| $\alpha_t$ | 0.0000 | 0.0101 | 0.0000 | 0.0010 | 0.0000 | 0.0000 |

\* p<0.1, ** p<0.05, *** p<0.01

Notes: Std. Err. clustered by sector (i.e. 2 digits NACE code) in parentheses; "Country FE" ("Sector FE" / "Year FE") shows the p-value associated with the Wald test of the joint null hypothesis that country (sector / year) fixed effects are not statistically significant. Overdispersion tests (p-values): 295.63 (0.0000), 268.21 (0.0000), 265.26 (0.0000). The regression also includes a dummy variable equal to one when there is a missing value in IFA.